\documentclass[aps,nofootinbib]{revtex4-1}

\usepackage{fancyhdr}
\usepackage{latexsym}
\usepackage{amsfonts}
\usepackage{tabularx}

\usepackage{float}
\usepackage{subfloat}

\usepackage{graphicx}
\usepackage{mathrsfs}
\usepackage{amssymb}
\usepackage{amsmath}
\usepackage{verbatim}
\usepackage{multirow}
\usepackage{color}
\usepackage{soul}
\usepackage{slashed}
\usepackage[colorlinks=false,
pdfborder={0 0 0}]{hyperref}

\newcommand{\TeV}{{\ensuremath\rm TeV}}
\newcommand{\GeV}{{\ensuremath\rm GeV}}
\newcommand{\MeV}{{\ensuremath\rm MeV}}

\newcommand{\eqn}{equation}

\newcommand{\lb}{\left(}
  \newcommand{\rb}{\right)}

\newcommand{\lam}{\lambda}

\def\D0{\slash\!\!\!\!\!\!\!\!\!\:D0}
\newcommand{\HS}{\texttt{HiggsSignals}}

\newcommand{\HSv}[1]{\texttt{HiggsSignals-#1}}
\newcommand{\HBv}[1]{\texttt{HiggsBounds-#1}}
\newcommand{\Micro}{\texttt{MicrOmegas}}
\newcommand{\nn}{\nonumber}

\newcommand{\oblique}{Altarelli:1990zd,Peskin:1990zt,Peskin:1991sw,Maksymyk:1993zm}

\newcommand{\lp}{\lambda_5}

\newcommand{\fr}{\frac}

\newcommand{\checkmate}{\texttt{CheckMATE}}
\newcommand{\mgfive}{\texttt{MG5\_aMC@NLO}}

\begin{document}
%\preprint{\parbox[t]{3.3cm}{DESY 18-222 }}

\bibliographystyle{hunsrt} %% includes arxiv numbers

\date{\today}

\title{Constraining the Inert Doublet Model using Vector Boson Fusion}
\vspace*{1.0truecm}
\author{Daniel Dercks\vspace{0.2cm}}
%\email{daniel.dercks@desy.de}
%\affiliation{DESY Hamburg, Notkestrasse 85, 22607 Hamburg, Hamburg, Germany\vspace{0.2cm}}
\author{Tania Robens}
%\email{trobens@irb.hr}
%\affiliation{Theoretical Physics Division, Rudjer Boskovic Institute, 10002 Zagreb, Croatia \vspace{0.5cm}}
\renewcommand{\abstractname}{\vspace{0.5cm} Abstract}

\begin{abstract}
  \vspace{0.5cm}
%\abstract{
In this work, we use a recast of the Run II search for invisible Higgs decays within Vector Boson Fusion to constrain the parameter space of the Inert Doublet model, a two Higgs doublet model with a dark matter candidate. When including all known theoretical as well as collider constraints, we find that the above can rule out a relatively large part in the $m_H,\,\lam_{345}$ parameter space, for dark scalar masses $m_H\,\leq\,100\,\GeV$. Including the latest dark matter constraints, a smaller part of parameter space remains which is solely excluded from the above analysis. We also discuss the sensitivity of monojet searches and multilepton final states from Run II.
\end{abstract}

%}

\maketitle

\maketitle

\newpage

\section{  Introduction}
\label{Sec:Intro}
The Inert Doublet model (IDM)  is one of the most straighforward extensions of the Standard Model (SM) \cite{Deshpande:1977rw,Cao:2007rm,Barbieri:2006dq}. It belongs to the class of Two Higgs Doublet Models (2HDM) which contain two $SU(2)$ doublets in the scalar sector. One of these doublets,  $\phi_S$, has a nonvanishing vacuum expectation value (vev) which is responsible for the spontaneous breaking of electroweak symmetry in the Standard Model while the second scalar doublet $\phi_D$ by construction does not acquire such a vev. This second doublet is hence not involved in the spontaneous mass generation in the Standard Model and does not couple to the SM fermions.

Within this model we impose an additional Z$_2$ symmetry, labelled $D$-symmetry, defined via the transformation

{
  \begin{equation}
    \phi_D \to - \phi_D, \,\,
    \phi_S\to \phi_S, \,\,
    \text{SM} \to \text{SM}, \label{eq:dsym}
  \end{equation}
}
which should be respected by the Lagrangian and  the vacuum.

As electroweak symmetry breaking in this model proceeds completely analogous to the SM without the second doublet, $\phi_S$ {provides} the SM-like Higgs particle and is assumed to be even under the $D$ symmetry. The second \emph{inert} or \emph{dark doublet} contains two charged and two neutral scalars and as they are odd under {the} imposed $D$-parity, its lightest neutral component provides a natural candidate for dark matter (DM). It provides a \lq\lq{}perfect example\rq\rq{}  of a WIMP \cite{LopezHonorez:2006gr,Honorez:2010re,Dolle:2009fn,Sokolowska:2011aa}, and leads to an interesting pattern for the evolution of the Universe, towards the Inert phase as given by the IDM, with one, two or three
phase transitions \cite{Ginzburg:2010wa}. Furthermore, the IDM can provide a strong first-order phase transition \cite{Hambye:2007vf,Chowdhury:2011ga,Borah:2012pu,Gil:2012ya,Blinov:2015vma} as required by the Sakharov conditions to generate a baryon asymmetry of the Universe.
After the discovery of a SM-like Higgs particle in 2012, many studies have been performed in the context of the IDM which use  Higgs measurements as well as astrophysical observations, {see e.g. {\cite{Swiezewska:2012eh,Gustafsson:2012aj,Arhrib:2012ia,Arhrib:2013ela,Krawczyk:2013jta,Goudelis:2013uca,Belanger:2015kga,Blinov:2015qva,Ilnicka:2015jba,Ilnicka:2018def}}}.\footnote{{Recent analyses for models which extend the IDM by an additional singlet have been performed in \cite{Banik:2014cfa,Banik:2014eda,Bonilla:2014xba,Plascencia:2015xwa}.}} {In addition,} proposals {were made} how to search for  dark scalars at {the} LHC in leptonic final states \cite{Dolle:2009ft,Swaczyna,Gustafsson:2012aj,Ilnicka:2015jba,Datta:2016nfz,Wan:2018eaz} and in single or dijet channels \cite{Belyaev:2018ext,Poulose:2016lvz}.

Recently, also the important issue of vacuum (meta-) stability in the IDM has been discussed, and it was found that additional, possibly heavy scalars can have a strong impact on it \cite{Kadastik:2011aa,Goudelis:2013uca,Swiezewska:2015paa,Khan:2015ipa}.\footnote{Similar solutions can be found in a simple singlet extension of the SM Higgs sector, cf. e.g. \cite{Pruna:2013bma,Robens:2015gla} and references therein.} 

While the model is intruiging per se and in spite of benchmark scenarios for the current LHC run \cite{Ilnicka:2015jba,deFlorian:2016spz}, it has not yet been studied explicitly by the LHC collaborations. However, recasts of other BSM searches with similar topologies have been presented in the literature, with prominent examples for searches for supersymmetric particles at LEP \cite{EspiritoSanto:2003by} as well as the first LHC run \cite{Belanger:2015kga}. \\

In this work, we present a recast of the Run II analyses presented in Ref.~\cite{Sirunyan:2018owy} by the CMS collaboration which target an invisibly decaying SM-like Higgs boson produced in vector boson fusion (VBF), and Ref.~\cite{ATLAS-CONF-2017-060} by the ATLAS collaboration which focusses on  monojet final states. We reinterpret the results of these searches within the IDM by making use of the \checkmate{} \cite{Drees:2013wra,Dercks:2016npn} framework. 

The regions considered in this work are tested against all currently available theoretical and experimental constraints, with scan procedure and limits as described in Refs.~\cite{Ilnicka:2015jba,Ilnicka:2018def,Kalinowski:2018ylg}.

 We explore the reach of the above searches for the model's parameter space and identify regions which cannot be excluded by any of the other tested constraints. Finally, we briefly comment on other experimental BSM searches at LHC Run II that could be used as recasts for the IDM and are expected to yield further constraints on its parameter space.

\section{ The model}
\label{sec:model}
\noindent
Imposing symmetry under the $D$-transformation given in Eq.~(\ref{eq:dsym}), the full scalar potential of the IDM is given by 
 \begin{eqnarray}\label{pot}
                                                               \lefteqn{V(\phi_S, \phi_D) =}  \nn \\ && -\fr{1}{2}\left[m_{11}^2(\phi_S^\dagger\phi_S)\!+\! m_{22}^2(\phi_D^\dagger\phi_D)\right] +
                                                               \fr{\lambda_1}{2}(\phi_S^\dagger\phi_S)^2\! 
                                                   \nn     \\&&       +\!\fr{\lambda_2}{2}(\phi_D^\dagger\phi_D)^2 +\!\lambda_3(\phi_S^\dagger\phi_S)(\phi_D^\dagger\phi_D)\!
                                                               \!+\!\lambda_4(\phi_S^\dagger\phi_D)(\phi_D^\dagger\phi_S) \nn \\&& +\fr{\lambda_5}{2}\left[(\phi_S^\dagger\phi_D)^2\!
                                                               +\!(\phi_D^\dagger\phi_S)^2\right].
\end{eqnarray}

In this formulation, all parameters are real (see e.g. \cite{Ginzburg:2010wa}). 

Depending on the signs and values of the individual parameters in $V(\phi_S, \phi_D)$, the minimisation conditions may result in different vacuum configurations where none, one or both vevs of $\phi_S$ or $\phi_D$ are non-vanishing. Within this work, we focus on the IDM realisation $\langle \phi_S \rangle \neq 0, \langle \phi_D \rangle = 0$, for which the decomposition around the vacuum state is given by
\begin{equation} \label{dekomp_pol}
  \phi_S = \begin{pmatrix}\phi^+\\ \frac{1}{\sqrt{2}} \lb v+h+i\xi \rb \end{pmatrix}, \phi_D  = 
  \begin{pmatrix} H^+ \\ \frac{1}{\sqrt{2}} \lb H+iA \rb  \end{pmatrix}.
\end{equation}
Here, $v\,=\,246\,\GeV$ denotes the SM vacuum expectation value and the scalar field component of $\phi_S$ contains the SM-like Higgs boson $h$ with mass
\begin{equation}\label{Higgsmass}
  m_{h}^2=\lambda_1v^2= m_{11}^2, 
\end{equation}
fixed by the experimentally observabed value of $125.1\,\GeV$.

In addition to the components known from the Standard Model, the second scalar doublet of the IDM, $\phi_D$, contains four \emph{dark} or \emph{inert} scalar field components $H,\,A,\,H^\pm$ with masses given as follows: 
\begin{align}
  m_{H^\pm}^2&=\fr{1}{2} \left(\lambda_3 v^2-m_{22}^2\right),\label{eq:m1} \\
  m_{A}^2&=m_{H^\pm}^2+\fr{1}{2}\left(\lambda_4-\lambda_5\right)v^2\ =\fr{1}{2}( \bar{\lambda}_{345}v^2-m_{22}^2), \label{eq:m2} \\
  m_{H}^2&=m_{H^\pm}^2+\fr{1}{2}\left(\lambda_4+\lambda_5\right)v^2\ = {\fr{1}{2}}(\lambda_{345}v^2-m_{22}^2 ), \label{eq:m3}
\end{align}
\noindent
where we have defined 
\begin{\eqn*}
\lambda_{345}:=\lambda_3 + \lambda_4 + \lambda_5; {\bar{\lambda}_{345}:=\lambda_3 + \lambda_4 - \lambda_5}.
\end{\eqn*} 
While their interactions with the Standard Model vector bosons can be derived from the gauge kinetic term in the Lagrangian, the absence of any gauge invariant Yukawa-like interaction between $\phi_D$ and the Standard Model fermion sector prohibits any tree level interactions between these four dark particles and the SM fermions. 
Moreover, due to the exact $D$-symmetry the lightest neutral scalar cannot decay and may therefore provide a candidate for dark matter.\footnote{Charged DM has been strongly limited by astrophysical analyses \cite{Chuzhoy:2008zy}.} Note that, contrarily to generic Two-Higgs-Doublet-Models which denote $H/A$ as the scalar/pseudoscalar components of a doublet, we cannot make such a unique idenfication here as there is no interaction of $\phi_D$ with the Standard Model fermions. In fact, we can swap the roles of $H$ and $A$ by making the replacement $\lam_5\,\leftrightarrow\,-\,\lam_5$, cf. Appendix \ref{app:fr}.

Within this work, we make the choice ${m_H} < m_A, m_{H^\pm}$ and assume $H$ to be the DM candidate. According to Eqs.~(\ref{eq:m1}-\ref{eq:m3}), this choice implies the relations  $\lp<0$ and ${{\lam_{45}}}:=\lambda_{4}+\lambda_{5}<0$. The parameters $\lambda_{345}$ and $\bar \lambda_{345}$ are related to {the} triple and quartic coupling  between the SM-like Higgs $h$ and the DM candidate $H$ or the scalar $A$, {respectively}. $\lambda_3$ is relevant for   
the $h$  interaction with the charged scalars $H^\pm$. Lastly, the parameter $\lambda_2$ describes the quartic self-couplings of dark particles. A list of all relevant Feynman rules for this model is provided in Appendix \ref{app:fr}.

Starting from the general scalar potential in Eq.~(\ref{pot}), the IDM has 7 degrees of freedom. As $\phi_S$ plays the same role as the SM Higgs doublet for electroweak symmetry breaking, the two parameters $m_h$ and $v$ are fixed by the Higgs mass measurement and electroweak precision data, respectively. We are therefore left with 5 degrees of freedom which we choose to be the physical parameters\begin{\eqn*} 
(m_H,m_A,m_{H^{\pm}}, \lam_2, \lam_{345} ). 
\end{\eqn*}
From these the corresponding dependent values of the other theory parameters can be derived by applying the relations given above. 

\section{ Constraints}\label{sec:constraints}
As has been widely discussed in the literature, the IDM is subject to numerous constraints which can be derived from both theoretical grounds as well as experimental results. We briefly remind the reader of these constraints here and refer to the literature \cite{Kanemura:1993hm,Akeroyd:2000wc,LopezHonorez:2006gr,Cao:2007rm,Gustafsson:2007pc,Tytgat:2007cv,Agrawal:2008xz,Lundstrom:2008ai,Dolle:2009fn,Dolle:2009ft,Arina:2009um,Honorez:2010re,Krawczyk:2009fb,Gustafsson:2010zz,Swiezewska:2012ej,Gustafsson:2012aj,Modak:2015uda,Ilnicka:2015jba,Hashemi:2015swh,Diaz:2015pyv,Ferreira:2015pfi,Blinov:2015qva,Longas:2015sxk,Hashemi:2016wup,Datta:2016nfz,Alves:2016bib,Kanemura:2016sos,Poulose:2016lvz,Borah:2017dfn,Eiteneuer:2017hoh,Ilnicka:2018def,Wan:2018eaz,Heisig:2018kfq,Belyaev:2018ext,Kalinowski:2018ylg} for further details. 

The constraints we use in this work have been extensively discussed in Refs.~\cite{Ilnicka:2015jba,Ilnicka:2018def,Kalinowski:2018ylg} and we refer the reader to {these references} for more detailed explanations. Here, we only summarise all relevant constraints and {point to updates on experimental limits whenever applicable}. The calculation of the IDM spectrum and tests of several of the below {bounds} have been obtained using \texttt{2HDMC} \cite{Eriksson:2010zzb}.

\subsection{ Theoretical constraints}\label{sec:thconst}

We apply the following theoretical constraints:

\begin{itemize}
\item The vacuum of the model needs to be bounded from below.\footnote{The {conditions} are applied at tree level; see e.g. Refs.~\cite{Goudelis:2013uca,Swiezewska:2015paa} for a discussion of changes using higher-order predictions.} These lead to the conditions
  \begin{align}
    \lambda_1 > 0, \lambda_2 > 0, \lambda_3 + \sqrt{\lambda_1 \lambda_2} > 0, \lambda_{345} + \sqrt{\lambda_1 \lambda_2} > 0
    \end{align}
\item All couplings  must allow for a perturbative discussion which is why we restrict all couplings to be smaller than $4 \pi$.
\item All {$2\,\rightarrow\,2$} scalar scattering processes must not violate perturbative unitarity and we apply standard bounds as implemented in \texttt{2HDMC}.

\item  In generic Two Higgs Doublet Models, several vacua can coexist. The tree level condition to be in the inert vacuum has been calculated in

   \cite{Ginzburg:2010wa,Gustafsson:2010zz,Swiezewska:2012ej}
  \begin{\eqn}\label{eq:invac}
    \frac{m_{11}^2}{\sqrt{\lam_1}}\,\geq\,\frac{m_{22}^2}{\sqrt{\lam_2}}
  \end{\eqn}

Here, $m_{11}$ and $m_{22}$ can directly be derived from Eqs.~(\ref{Higgsmass}) - (\ref{eq:m3}) and the above condition translates to
  \begin{\eqn*}
    \lam_{345}\,\leq\,\frac{\sqrt{\lam_2}\,m_h\,v+2 m_H^2}{v^2}
  \end{\eqn*}
The above constraint links the value of $\lam_{345}$ and the dark scalar mass to the coupling $\lam_2$ which describes self-couplings in the scalar sector and has no influence on collider phenomenology (see e.g.\ the discussion in \cite{Kalinowski:2018ylg}). Requiring the Higgs self-coupling vertices to acquire maximally allowed values of $4\,\pi$ leads e.g.\ to $\lam_2\,\lesssim\,4$ \cite{Ilnicka:2015jba}. This bound, in the parameter region with
relatively light dark scalars with masses $m_H\,\lesssim\,100\,\GeV$, would result in $\lam_{345}\,\leq\,\mathcal{O}(1)$. Note, however, that this bound is not completely mandatory. Several minima may coexist (see e.g.~\cite{Ginzburg:2007jn,Sokolowska:2011yi}) and the inert one may only be a local one as long as the transition time to the global non-inert minimum is sufficiently large. Moreover, the above condition may be significantly altered at next-to-leading order, see e.g.\ Refs.~\cite{Castillo:2015kba,Ferreira:2015pfi}. The next-to-leading order effects are however quite involved and can not easily be generalized, but need to be recalculated on a case-by-case basis \footnote{We thank P. Ferreira and B. Swiezewska for useful discussions regarding this point.}. In this work, we focus on current constraints from LHC searches that are independent of $\lam_2$; we therefore also consider values of $\lam_{345}\,\gtrsim\,1$. In case of a discovery, a detailed analysis would be needed in order to correctly evaluate the above condition beyond leading order, see e.g.\ related studies in Refs.~\cite{Castillo:2015kba,Lindner:2016kqk}.
\end{itemize}

\subsection{Experimental constraints}\label{sec:expconst}
In addition to the theoretical {bounds} listed above, several experimental observations put tight constraints on the parameter space of the IDM:
\begin{itemize}
\item We fix the mass of the SM-like Higgs boson $h$ to 
\begin{align}
m_h=125.1\,\GeV \label{Eq:mhexp}
\end{align} in agreement with the results from the LHC experiments \cite{Aad:2015zhl}. Note that this has already been accounted for when we chose $(m_H,m_A,m_{H^{\pm}}, \lam_2, \lam_{345} )$ as the five degrees of freedom of the IDM.

\item We furthermore require the {total} width of the 125\,\GeV{} Higgs to obey Ref.~\cite{CMS-PAS-HIG-18-002}
  \begin{\eqn*}
    \Gamma_\text{tot}\,\leq\,9\,\MeV
  \end{\eqn*}
  which {is applicable in} those regions of parameter space which predict additional decays of the SM-like Higgs boson.
\item
  Furthermore, we take into account strong bounds from the measured total widths of the electroweak SM gauge bosons, cf. e.g. Ref.~\cite{Agashe:2014kda}, by forbidding potentially dangerous kinematic mass configurations via the following hard constraints:

  \begin{align}\label{eq:gwgz}
    m_{A,H}+m_{H^\pm} &\geq m_W, \\
    m_A+m_H &\geq m_Z, \\
    2 m_{H^\pm} &\geq m_Z.
  \end{align} 

\item We furthermore require a $2\,\sigma$, i.e. $95 \%$ C.L., agreement with electroweak precision observables, parameterized through the electroweak oblique parameters $S,T$ and $U$ \cite{\oblique}.

\item In order to evade bounds from long-lived charged particle searches, we conservatively set an upper limit on the charged scalar lifetime of $\tau\,\leq\,10^{-12}\,s$, to guarantee decay before the innermost detector layer. This translates to a lower bound on the total decay width of the charged scalar $H^\pm$ of $\Gamma_\text{tot}\,\geq\,6.58\,\times\,10^{-13}\,\GeV$. Mass dependent bounds on the charged scalar lifetime have been studied in detail in Ref.~\cite{Heisig:2018kfq}.

\item
  A bound on the lower mass of $m_{H^\pm}$ has been derived in Ref.~\cite{Pierce:2007ut}. Although a more dedicated analysis of this bound within the current models' framework would be required, 
  we take $m_{H^\pm}\,\geq\,70\,\GeV$ as a conservative lower limit.
\item We also {require agreement with} the  null-searches from the LEP, Tevatron, and LHC experiments using \\{\HBv5.2.0beta}~\cite{Bechtle:2008jh,Bechtle:2011sb,Bechtle:2013wla,Bechtle:2015pma}, including all experimental bounds up to Moriond 2017.\footnote{Please see the tool's documentation material in Ref.~\cite{hbwebpage} for a detailed discussion of the included limits.}
\item We update the limits on the invisible decay of $m_h$ and take the results presented in Ref.~\cite{Khachatryan:2016whc} which require $\text{BR}_{h\,\rightarrow\,\text{inv}}\,\leq\,0.24$.
\item Furthermore, we apply new limits on the branching ratio $h\,\rightarrow\,\gamma\,\gamma$ taken from \cite{Khachatryan:2016vau} and require $\mu\,=\,1.14^{+0.19}_{-0.18}$. Since within the IDM the production cross sections of the SM-like Higgs are unaffected, we use the bound on $\mu$ in combination with the Standard Model value \cite{deFlorian:2016spz} of $\text{BR}\,(h\,\rightarrow\,\gamma\,\gamma)\,=\,2.270\,\times\,10^{-3}$ and require
  \begin{\eqn}\label{eq:rgaganew}
    \text{BR}\,(h\,\rightarrow\,\gamma\,\gamma)\,\in\,\left[1.77;3.45\right]\times\,10^{-3}
  \end{\eqn}
  at the two-sigma level. 
\item{
    In addition, we require agreement within $2\,\sigma$ for the 125 GeV~Higgs signal strength measurements. For this, we make use of the publicly available tool\\ {\HSv2.2.1beta}~\cite{Bechtle:2013xfa}, {and require $\Delta \chi^2\,\leq\,11.3139$}, corresponding to the $95 \%$ confidence level of a 5-dimensional fit.\footnote{{We used a combination of Run I combination, Run II, and simplified template cross sections} within \HS.}} 
\item
  We also include limits on the model's parameter space that have been obtained in previous reinterpretations of collider dark matter searches, predominantly within supersymmetric scenarios. Major limits stem from the reinterpretation of a LEP analysis \cite{EspiritoSanto:2003by} within the IDM framework \cite{Lundstrom:2008ai}. This particularly rules out all regions where
  \begin{align}
    m_A &\leq 100\,\GeV, \\
    m_H &\leq 80\,\GeV, \\
    \Delta m {(A,H)} &\geq\,8\,\GeV
  \end{align}
  are simultaneously fulfilled.
\item{}
  After taking into account all the above limits we are outside of the region excluded due to the recent reinterpretation of the SUSY analysis from LHC Run I \cite{Belanger:2015kga}. 

\item{} We apply dark matter relic density limits obtained by the Planck experiment \cite{Aghanim:2018eyx}:
  \begin{\eqn}\label{eq:planck}
    \Omega_c\,h^2\,=\,0.1200\,\pm\,0.0012
  \end{\eqn}
  In this work, we do not require the dark matter candidate of the IDM to provide the full relic density, but use it as an upper limit\footnote{{In such a scenario, additional dark matter candidates would be needed in order to account for the missing relic density; {cf. e.g. Ref.~\cite{Cheung:2012qy} for a dedicated discussion of such scenarios within a supersymmetric setup.}}}. 
  Being conservative, we require
  \begin{\eqn}\label{eq:planck_up}
    \Omega_c\,h^2\,\leq\, 0.1224,
  \end{\eqn}
  which corresponds to not overclosing the universe at $95\,\%$ confidence level. In addition to this bound,  we specifically identify those regions which reproduce the observed DM density within the 2 $\sigma$ interval around the above best fit value value. The dark matter relic density has been calculated using \Micro{} version 4.3.5 \cite{Barducci:2016pcb}.
\item{} Regarding direct detection dark matter constraints, we compare to the most recent results of XENON1T \cite{Aprile:2018dbl}.\footnote{We here use the data available from Ref.~\cite{PhenoData} in a digitalized format. In our code, we use an approximation function which reproduces these constraints on the per-cent level.}

  As before, we consider the possibility of a multi- component dark matter scenario in which the IDM only makes up for a fraction of the total dark matter relic density. In this case, the upper limit from direct detection depends on the actual DM relic density for the specific point in parameter space; therefore, we have to introduce a rescaling factor, leading to the (relic density dependent) limit
  \begin{\eqn}\label{eq:xenonlim}
    \sigma\,(m_H, \{\ldots\})\,\leq\,\sigma^\text{XENON1T}(m_H) \times\,\frac{\Omega^\text{Planck}}{\Omega (m_H, \{\ldots\})}, 
  \end{\eqn}
  where $m_H$ now denotes the dependence on the mass of our dark matter candidate $H$ and $\{\ldots\}$ is short for all other parameters specifying the respective IDM parameter point.\footnote{{See also Refs.~\cite{Cao:2007fy,Cheung:2012qy,Belanger:2014bga,Belanger:2014vza,Badziak:2015qca}.}} Direct detection cross sections are again obtained using \Micro.
\end{itemize}

The scan setup has been described in great detail in Ref.~\cite{Ilnicka:2015jba}. To determine allowed regions in parameter space, we follow the procedure discussed therein, including the experimental updates listed above.

\section{LHC Analysis of VBF and Monojets}

In this work, we choose to constrain ourselves to cases for dark matter candidate masses $m_H\,\leq\,100\,\GeV$. Due to the relatively high production cross section in such cases, these will be the regions which are most sensitive to collider searches (see e.g. \cite{Poulose:2016lvz,Belyaev:2016lok,Datta:2016nfz,Belyaev:2018ext} for recent work on low mass scenario studies at the LHC). 

We here concentrate on the 13 \TeV{} CMS search for an invisibly decaying Higgs  \cite{Sirunyan:2018owy} produced through vector boson fusion (VBF) and a 13 \TeV{} ATLAS search \cite{ATLAS-CONF-2017-060} for dark matter candidates in the monojet channel. These respectively lead to the collider signatures
\begin{align}\label{eq:sig}
  p\,p\,&\rightarrow\,j\,j\,+\,\slashed{E}_T, \quad \text{(VBF)}\\
  p\,p\,&\rightarrow\,j\,+\slashed{E}_T, \quad \text{(Monojet)}.
\end{align} 
In this study, we mainly focus on the VBF channel which, as we show later, provides the strongest sensitivity. We however also determine bounds on the IDM using a monojet reinterpretation for comparision. A dedicated exploration of this channel including sensitivity prospects of the high luminosity LHC can be found in Ref.~\cite{Belyaev:2018ext}.
\subsection{Features of the VBF Channel}
The two jets in the VBF channel typically have a large separation in pseudorapidity. The corresponding cuts used in the above analysis are listed in table \ref{tab:selections2}. These form a ``Cut-and-Count analysis'' and a ``Shape analysis''. The former is designed for a large signal-to-background ratio and requires a large value for the invariant mass $m_{jj}$ of the jet pair, whilst the latter defines several signal regions binned in $m_{jj}$ and used collectively in a fit. Using these signatures, the CMS collaboration finds an upper limit on the invisible Higgs branching ratio of $\text{BR}_{h\,\rightarrow\,\text{inv}}^\text{max}\,=\,0.53$ using the cut-and-count analysis and $0.28$ for the shape analysis which are both weaker than the upper limit $\text{BR}_{h\,\rightarrow\,\text{inv}}^\text{max}\,=\,0.24$ used as a hard cut in our scan (see section \ref{sec:expconst}). {This constraint is only applicable to parameter points in the IDM for which $m_H < m_h/2$. However, points with heavier scalars would also predict additional signal events in the above analysis due to processes with off-shell $h$ production ($p p \rightarrow h^* j j\rightarrow H H j j$) and contributions from decay chains with hadronically decaying final state particles (e.g.\ $p p \rightarrow H^\pm H \rightarrow j j H H$). We \emph{recast} the above mentioned VBF analysis in the context of these processes to potentially extract additional constraints applicable to regions with larger values of $m_H$.}

\newcolumntype{Y}{>{\centering\arraybackslash}X}
\begin{table}
  \begin{tabular}{lcc}
    \hline
    \hline
    Requirement                                     & Cut-and-Count                           & Shape          \\
    \hline
    Leading Jet $p_{\rm T}$                 & \multicolumn{2}{c}{ $>$80~GeV                             }            \\
    Second Jet $p_{\rm T}$                 & \multicolumn{2}{c}{$>$40~GeV} \\
    $E_{\rm T}^{\rm miss}$                  & \multicolumn{2}{c}{$>$250~GeV}\\
    $|\Delta\phi_{j,E_{\rm T}^{\rm miss}}|$ & \multicolumn{2}{c}{$>$0.5}  \\
    $|\Delta\phi_{jj}|$                     & \multicolumn{2}{c}{$<$1.5}          \\
    $\eta_1 \cdot \eta_2$ & \multicolumn{2}{c}{$<$0} \\
    $|\Delta\eta_{jj}|$                     & $>$4.0                                &  $>$1.0    \\
    $m_{jj}$                                & $>$1.3~TeV              & $>$ 200~GeV (binned) \\
    \hline
    \hline
  \end{tabular}
  \caption{\label{tab:selections2} Summary of the main kinematic requirements in the signal regions in Ref.~\cite{Sirunyan:2018owy}.}
\end{table}

\subsection{Simulation and Validation of the VBF Channel}
\label{sec:sim}
In this work, we concentrate on the above VBF search which has been implemented within the \checkmate{} \cite{Drees:2013wra,Dercks:2016npn} framework. \checkmate{} uses simulated event files for any BSM model, applies detector efficiencies and follows the event selection procedure of the implemented BSM searches from ATLAS and CMS to determine if any resulting signal prediction would violate the corresponding experimental bound.\footnote{For more information about how \checkmate{} works we refer to the corresponding manuals in Refs.~\cite{Drees:2013wra,Dercks:2016npn}.  We implemented the above mentioned VBF search using the \texttt{AnalysisManager} tool described in Ref.~\cite{Kim:2015wza}.} Validation has been performed {by reproducing the quoted numbers expected from the Standard Model Higgs boson with 100\% invisible branching ratio. Following the procedure described in the experimental publication, we use} the POWHEG-Box \cite{Nason:2004rx,Frixione:2007vw,Nason:2009ai,Alioli:2010xd} for simulating Monte-Carlo events at next-to-leading order in QCD and subsequently interface it to \texttt{Pythia 6.4.21} \cite{Sjostrand:2006za} to account for parton showering and hadronization of the final state. We perform the simulation separately for vector-boson-fusion (vbf) and gluon-initiated final states (ggf) which may also pass the aforementioned cuts. 

{As we are bound to leading order Monte Carlo tools for the simulation of the IDM, we additionally generate tree-level parton events with \mgfive{} \cite{Alwall:2011uj} showered with \texttt{Pythia 8.219} \cite{Sjostrand:2014zea} --- the same tools which we use for our subsequent IDM analysis --- to quantify the effect of an LO-only simulation. Both event samples are processed with \checkmate{} and the resulting signal predictions are shown in table \ref{tbl:validation}}.

\begin{table*}
  \begin{tabularx}{\textwidth}{ l | Y Y | Y Y Y Y }
    \hline \hline
    Region & Data & Background & \multicolumn{4}{c}{SM prediction with BR($h \rightarrow $invisible) = 100 \%} \\
    &  \multicolumn{2}{c|}{CMS} & CMS &  \multicolumn{3}{c}{Our Simulation}\\
    &  \multicolumn{2}{c|}{} &  & \texttt{Powheg-Box} & \texttt{MG5\_aMC@NLO} & Ratio\\
    \hline
    CutandCount & 2053 & 1779 $\pm$ 96 & 851 $\pm$	148 & 758  & 468 & 1.6\\
    $m_{jj} \in [200, 400]$~GeV & 16177 &14878 $\pm$ 566 & 591	$\pm$ 285 & 708 & 390 & 1.8 \\
    $m_{jj} \in [400, 600]$~GeV  &10008 & 9401 $\pm$ 387 & 571 $\pm$ 232 & 664 & 374 & 1.8 \\
    $m_{jj} \in [600, 900]$~GeV  &7277& 6658 $\pm$ 271 & 566	$\pm$ 172 & 737 & 433 & 1.7 \\
    $m_{jj} \in [900, 1200]$~GeV  &3138 & 2994 $\pm$ 144 & 472 $\pm$	131 & 483 & 293 & 1.7\\
    $m_{jj} \in [1200, 1500]$~GeV &1439 & 1283 $\pm$ 69& 307 $\pm$	64 & 314 & 202 & 1.7 \\
    $m_{jj} \in [1500, 2000]$~GeV &911 & 834 $\pm$ 51 & 344 $\pm$	83 & 319 & 203 & 1.6 \\
    $m_{jj} \in [2000, 2750]$~GeV &408 & 358 $\pm$ 29& 228 $\pm$	40 & 218 & 126 & 1.8 \\
    $m_{jj} \in [2750, 3500]$~GeV &87 & 73.8 $\pm$ 9.4& 90.3 $\pm$ 18.8 & 80.1 & 48.8 & 1.7\\
    $m_{jj} > 3500$~GeV           &29 & 30.3 $\pm$ 7.4& 37.4 $\pm$	9.1 & 38.2 & 19.9 & 1.9  \\
    \hline
    \hline
  \end{tabularx}
  \caption{Observed and expected number of events for all regions listed in Table \ref{tab:selections2}. SM predictions are determined for an entirely invisibly decaying Standard Model Higgs boson with $m_h = 125$~GeV produced both in  Vector Boson Fusion and Gluon Fusion. CMS numbers are taken from  Ref.~\cite{Sirunyan:2018owy} and compared to our numbers determined with  our analysis implementation in \checkmate{}, using both the LO-QCD generator \texttt{MG5\_aMC@NLO}  and the NLO-QCD Monte Carlo simulation \texttt{Powheg-Box}. Uncertainties quoted for CMS include both statistical and systematical uncertainties.}
  \label{tbl:validation}
\end{table*}

As can be seen, our setup reproduces the experimentally quoted results sufficiently well within the experimentally quoted error margin when using simulated events generated with an NLO-QCD Monte Carlo event generator. A leading-order Monte Carlo analysis, in comparision, significantly underestimates the signal prediction with a nearly constant ratio of $\approx 1.7$ across all signal regions. 

\begin{figure}
  \includegraphics[width=0.48\textwidth]{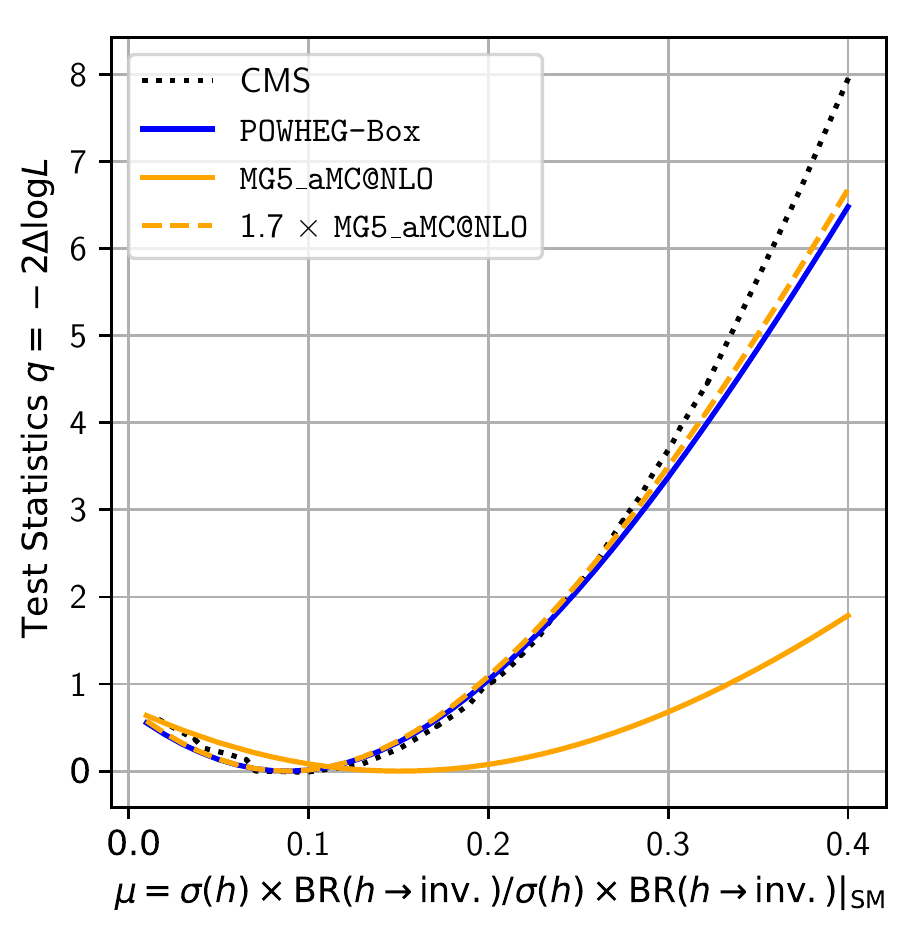}
  \caption{Comparision of the log-likelihood-ratio, using the numbers in Tab.~\ref{tbl:validation}.}
  \label{fig:likelihood}
\end{figure}

The numbers in Tab.~\ref{tbl:validation} can be used to derive upper limits on the invisible branching ratio of the Standard Model Higgs. For this purpose we employ a profile likelihood ratio test paired with the CLs prescription. For the shape analysis, we make use of the full background covariance matrix provided in Ref.~\cite{Sirunyan:2018owy}. As no such detailed information is provided for the signal, we conservatively assume that it is fully correlated across all bins. Our resulting distribution for the test statistics is shown in Fig.~\ref{fig:likelihood}. We are able to reproduce this distribution sufficiently well by either using our results determined with \texttt{POWHEG} or by rescaling the results of our leading-order simulation with a constant $K$-factor of $1.7$.

From our validation and the comparision of the above results using LO and NLO simulation, we conclude that a leading order simulation of the IDM signal is expected to systematically underestimate the correct number. However, a full next-to-leading order simulation of the off-shell VBF channel within the IDM is beyond the scope of this work. We determine results {for the IDM at leading order}. To be more precise, we simulate 
\begin{\eqn}\label{eq:vbfsig}
p p \rightarrow H H j j
\end{\eqn} 
event samples for the IDM within \texttt{MG5\_aMC@NLO} and \texttt{Pythia 8.219} by making use of the UFO model description of Ref.~\cite{Goudelis:2013uca}. {We here do not specify intermediate states, i.e. on parton level all processes leading to the final state given in Eqn. (\ref{eq:vbfsig}) have been taken into account. Besides VBF production of a SM-like scalar with invisible decays, this also includes e.g. $H\,A$ or $H^\pm\,H$ pair-production with the subsequent decay $ A\,\rightarrow\,H\,j\,j,\,H^\pm\,\rightarrow\,H\,j\,j$ or {vector boson scattering processes with} {dark scalars} in the $t-$channel. {Relative contributions of the latter to the total cross section become sizeable as $|\lam_{345}|\,\rightarrow\,0$.}} We include an invariant mass cut $m_{jj} \geq 130~\GeV$ and a pseudorapidity difference cut $\Delta \eta_{jj} \geq 0.5$ with $\eta_1 \eta_2 < 0$ in our parton event generation. Note that these are weaker than the signal region cuts in Tab.~\ref{tab:selections2}.

From our above findings, we expect our resulting bounds to be conservative. However, motivated from the results in Table \ref{tbl:validation} we also discuss the {limits we obtain} if our signal prediction is upscaled with the global $K$-factor of 1.7 motivated before to illustrate the potential impact of next-to-leading-order QCD effects.

\subsection{Features and Setup of the Monojet analysis}
\label{sec:submono}
Nearly any particle model with a dark matter candidate $H$ predicts the standard monojet signature $p p \rightarrow H H j$ for the LHC where the jet may originate from initial state radiation or, in some specific models other than ours, from the hard vertex. It is therefore to be expected that this channel is sensitive to the IDM in which $H$ plays the role of the dark matter candidate. A detailed analysis of this channel can be found in Ref.~\cite{Belyaev:2018ext}. However, as has for example been shown in Ref. \cite{Dutta:2017lny} in the context of a different model with similar topology, the vector boson fusion channel is expected to be significantly more sensitive than the monojet search. We reproduce this finding later. 

The analysis of the monojet channel is performed within \checkmate{}, similarly to the analysis above. As this analysis had already been implemented in the public code, we do not provide a separate validation here.\footnote{Validation material for this analysis can be found on the official \checkmate{} website, \url{https://checkmate.hepforge.org/AnalysesList/ATLAS_13TeV.html}.} We simulate the partonic process $q \bar{q}, g g \rightarrow H H j$ with \mgfive{} and apply a $p_T$ cut of 200 GeV on the leading jet, in accordance with the signal region requirement $p_T^j \geq $ 250 GeV of this analysis.

\section{Parameter space constraints}

In this section, we present the constraints resulting from the our recast of the searches for an invisibly decaying Higgs in both the vector boson fusion and the monojet channel. We initially consider parameter points which have passed all bounds presented in section \ref{sec:constraints}, {\sl apart from} the constraints imposed by dark matter bounds, i.e. dark matter relic density as well as direct detection, cf.\ Eqs.\ (\ref{eq:planck_up}) and (\ref{eq:xenonlim}). This approach allows for an investigation of the complementary between astrophysical and collider searches for this model.

\subsubsection*{Collider Constraints}

\begin{figure}
  \begin{minipage}{0.48\textwidth}
    \includegraphics[width=\textwidth]{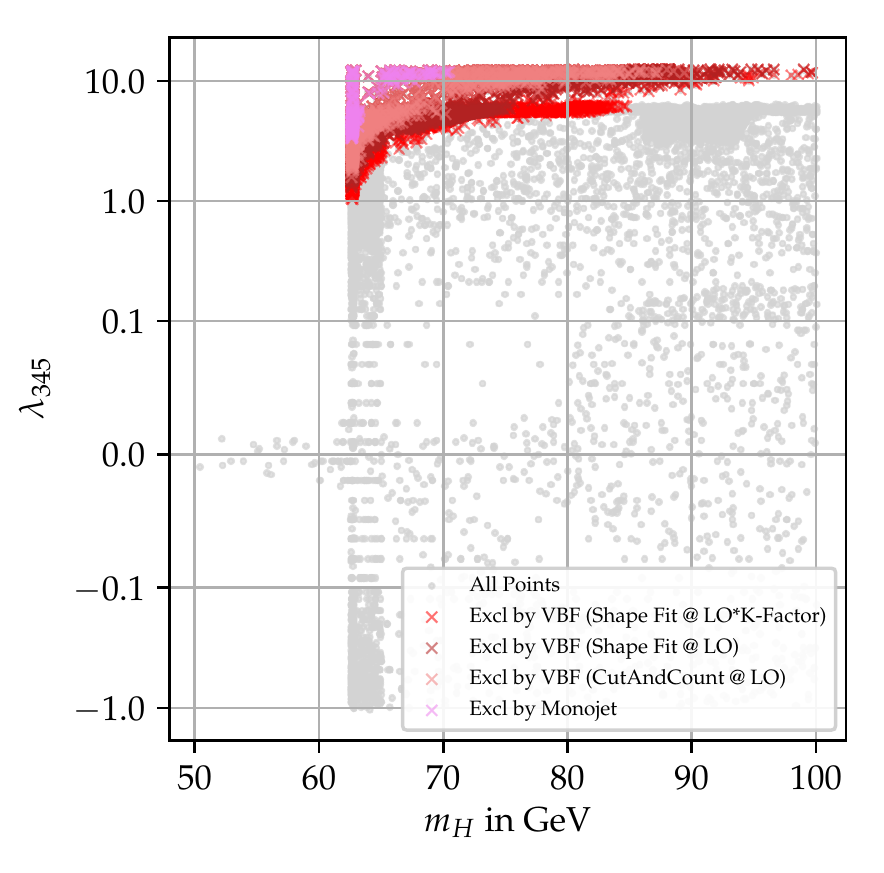}
  \end{minipage}
  \begin{minipage}{0.48\textwidth}
    \includegraphics[width=\textwidth]{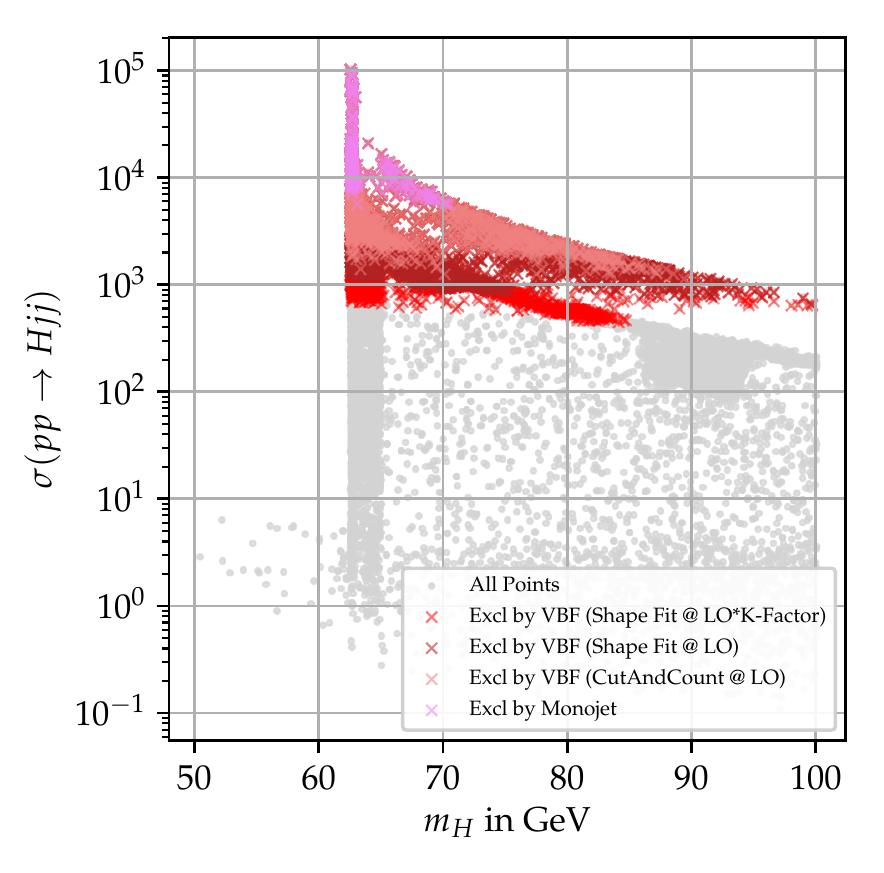}
  \end{minipage}
  \caption{\label{fig:nodmrecast} Allowed and excluded points after consideration of VBF and monojet analysis but without dark matter relic density and direct detection constraints. VBF results are shown using different {categories}, see in-text discussion. {\sl Left:} Results in $(m_H,\,\lam_{345})$ parameter plane. {\sl Right:} Results in $m_H$-$\sigma$ plane where $\sigma$ is the LHC VBF production cross section at 13 \TeV{} including the partonic cuts given in Sec.~\ref{sec:sim}.}
\end{figure}

We now demonstrate the effect of including the searches in Refs.~\cite{Sirunyan:2018owy}, \cite{ATLAS-CONF-2017-060} as introduced in the previous section. Our results are shown in Fig.~\ref{fig:nodmrecast} (left) where we only consider points which pass all prior constraints discussed in Sec.~\ref{sec:constraints}.\footnote{We note that the density of points has no theoretical meaning but is just a reflection of a bias in the generation of theoretical parameter tuples.} The general influence of these constraints has been discussed in detail in Refs.~\cite{Ilnicka:2015jba,Kalinowski:2018ylg} and will not be repeated here. For values $m_H\,\leq\,m_h/2$, it is especially the branching ratio limit on $h\,\rightarrow\,\text{invisible}$ which leads to the tight constraint $|\lam_{345}|\,\lesssim\,0.03$. For larger $m_H$ values, however,  $\lam_{345}$ can reach values up to the perturbativity limit $4\,\pi$ which has been imposed as a hard upper cut in the scan setup. Note that we have explicitly verified that the small stripe for $m_H > 80$ \GeV, $\lam_{345} > 6$ contains no viable parameter points as it is excluded by combining perturbativity requirements with limits on the electroweak oblique parameters and $R_{\gamma \gamma}$, see e.g.\ discussion in Ref.~\cite{Ilnicka:2015jba}.

We separately indicate which points are respectively excluded by the monojet and by the VBF search. For the latter, we explicitly distinguish the following exclusion categories.
\begin{itemize}
\item In \texttt{CutAndCount@LO}, we only determine the number of signal events in the cut-and-count signal region of Ref.~\cite{Sirunyan:2018owy} and use a single-bin likelihood ratio test to determine whether it is compatible with the numbers of observed and expected Standard Model events, see Tab.~\ref{tbl:validation}.
\item In \texttt{Shape Fit@LO} we determine the number of signal events in all $m_{jj}$ binned signal regions of {Tab.}~\ref{tbl:validation} and use a joint likelihood, including the background correlation matrix provided in Ref.~\cite{Sirunyan:2018owy}, to determine the overall $p$-value.
\item Whilst for the above two approaches we use the signal numbers as determined with the Monte Carlo generator \mgfive{} at leading order, for \texttt{Shape Fit@LO*K- Factor} we multiply all numbers with the constant $K$-factor of 1.7, c.f.\ discussion in Sec.\ref{sec:sim}.
\end{itemize}
According to our SM validation, we expect \texttt{LO} results to significantly underestimate the number of signal events and therefore lead to conservative bounds. Showing the results including the  $K$-factor determined from our SM validation renders an estimate of the impact of higher-order QCD contributions.

In general, we observe that a significant fraction of points can be constrained by the two collider searches considered in this work. As foreseen in Sec.~\ref{sec:submono}, the monojet channel shows a significantly reduced sensitivity as compared to the VBF search.\footnote{Our results appear to be compatible with former monojet sensitivity studies shown in Ref.~\cite{Belyaev:2018ext} which show no sensitivity for a benchmark point with $\lam_{345} = 1.7$ using an older version of the monojet search with only 10\% of the integrated luminosity that our analysis uses.} Though both channels suffer largely from SM QCD background sources, the VBF channel can make more precise predictions on the extected kinematics of the jets in the final state. This ultimatively allows for a higher signal-to-background ratio in the signal bins and thus results in a better sensitivity for many models in which both channels are present simultaneously. 

Whilst monojet studies alone are sensitive to values of $\lam_{345}$ down to 2.5 and $m_H$ masses in the range [$m_h/2$ - 70] GeV, we observe VBF reinterpretations to constrain $\lam_{345}$ down to 1 and extend the sensitivity range on $m_H$ values up to the maximum of 100~\GeV{} we consider. No parameter point with $m_H < m_h/2$ can be constrained as the small values of $\lam_{345}$ predict a far too small cross section. This can also be seen on the right of Fig.~\ref{fig:nodmrecast} where we show our bounds in terms of the VBF cross section, including the partonic cuts described in Sec.~\ref{sec:sim}. In fact, this region is largely constrained by the cut on $\text{BR}_{h\,\rightarrow\,\text{inv}}\,\leq\,0.24$ which we discuss in Sec.~\ref{sec:expconst}. As the VBF channel consists of one sub-measurement of this observable, it is evident that it cannot provide additional, stronger bounds than the one on the invisible branching ratio which has been used to generate our parameter samples.

Note that, though the bound is clearly very much dependent on the size of the cross section, we observe on the right of Fig.~\ref{fig:nodmrecast} that it is not flat in the $m_H$-$\sigma$-plane. This can be explained by differences in the signal efficiency from additional, small IDM contributions like $p p \rightarrow H^\pm H, H^\pm \rightarrow j j H$ which, in addition to $m_H$, also depend on the masses of the other inert scalars. The fact that the bound is not only dependent on the total cross section shows the importance of dedicated Monte Carlo recast analyses including off-shell effects.

\subsubsection*{Dark Matter Constraints}

\begin{figure}[tb!]
  \begin{minipage}{0.48\textwidth}
    \includegraphics[width=\textwidth]{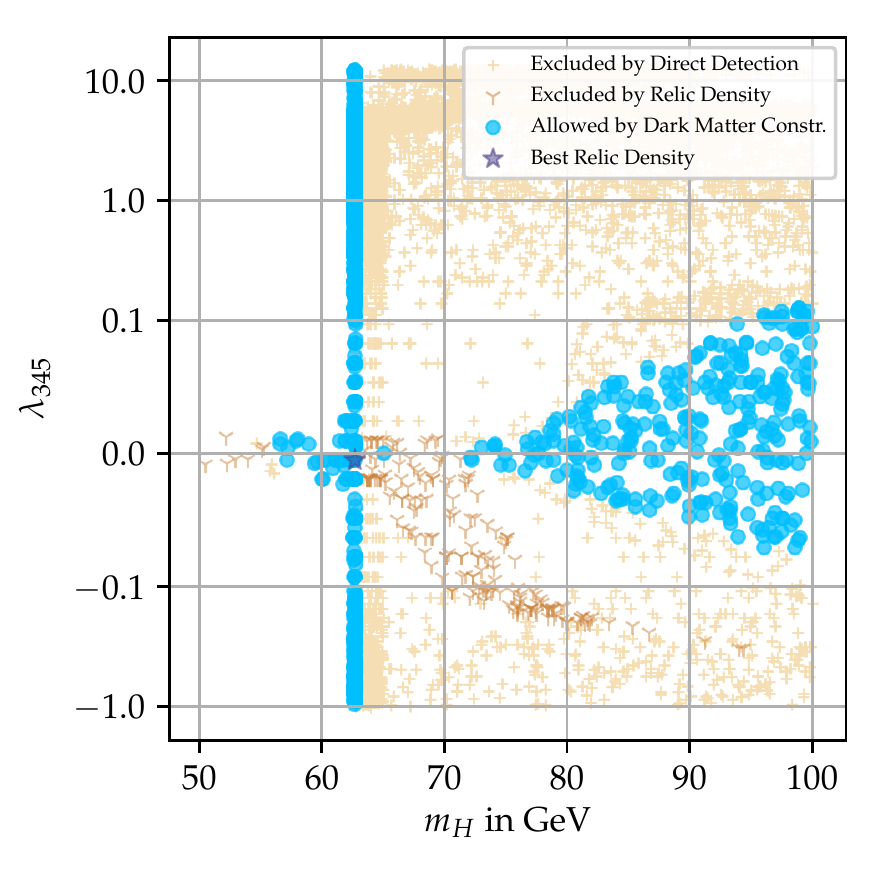}
  \end{minipage}
    \begin{minipage}{0.48\textwidth}
      \includegraphics[width=\textwidth]{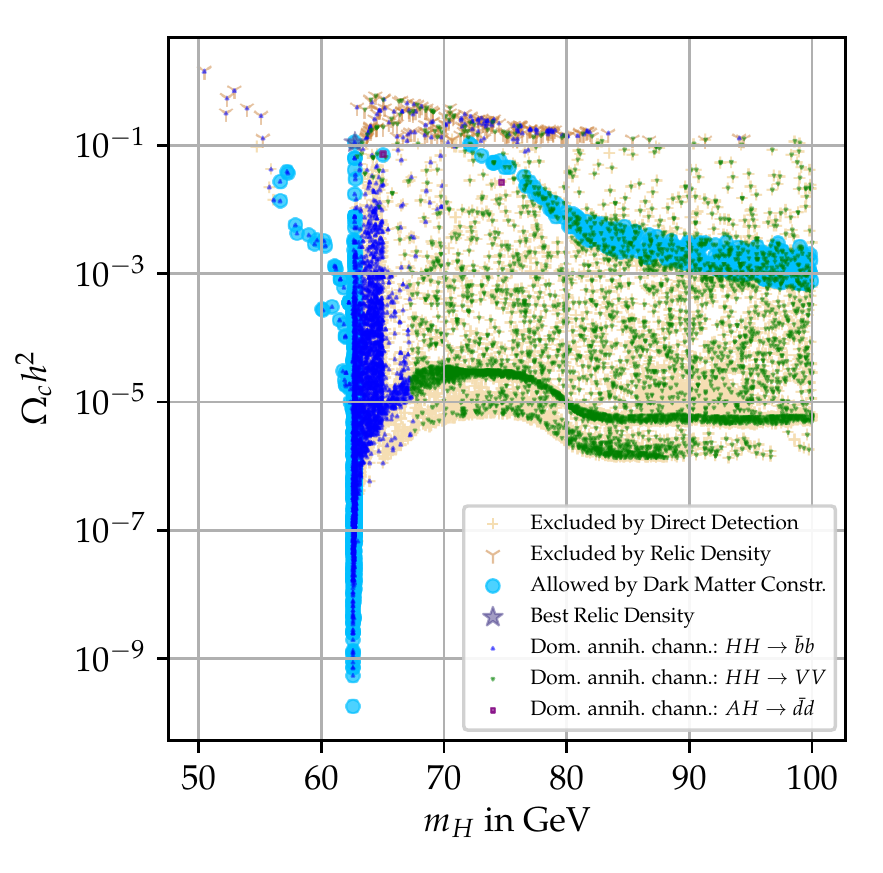}
    \end{minipage}
  \caption{\label{fig:dmconst} Parameter space after including dark matter relic density and direct detection constraints. {\sl Left:} allowed and forbidden regions in the $(m_H,\,\lam_{345})$ plane. {\sl Right:} Constraints in the $(m_H, \Omega_c h^2)$ plane. {On the right plot, we also show the dominant annihilation cross section for each parameter point.} The ``Best Relic Density'' point yields $\Omega_c h^2 = 0.1141$ which is the closest to the nominal Planck value, c.f. Eq.~(\ref{eq:planck}), out of all tested points. }
\end{figure}

We now impose the dark matter constraints specified by Eqs.~(\ref{eq:planck_up}) and (\ref{eq:xenonlim}) on the parameter space. As has been noted in Refs.~\cite{Ilnicka:2018def,Kalinowski:2018ylg}, it is especially direct detection constraints which have  improved by an order of magnitude with respect to the previous study in Ref.~\cite{Ilnicka:2015jba} which used the 2013 LUX results, c.f.\ Ref.~\cite{Akerib:2013tjd}. The parameter space is severely constrained, as is demonstrated in Fig.~\ref{fig:dmconst} where we now discuss the dark matter bounds on our parameter space without applying the VBF/Monojet limits. Fig.~\ref{fig:dmconst}, left, shows the results in the $m_H$-$\lam_{345}$ plane and Fig.~\ref{fig:dmconst}, right, displays the relic density abundance $\Omega_c h^2$ in dependence on $m_H$. {The second figure also labels the dominant annihilation channel for each tested parameter point as determined via \Micro.\footnote{For relatively small mass differences between the two dark neutral scalars $A$ and $H$, typically of a few \GeV, the co-annihilation channel $A\,H\,\rightarrow\,d\,\bar{d}$ becomes dominant. As this requires a relatively fine-tuned scenario, our scan only tested 2 such points. See also the discussion in \cite{Ilnicka:2015jba,Belyaev:2016lok}.}} We also indicate the point in our sample whose predicted value of $\Omega_c h^2 = 0.1141$ is closest to the Planck value in Eq.~(\ref{eq:planck}). This point yields 95~\% of the required cold dark matter relic density.  Especially for masses $m_H\,\geq\,63\,\GeV$, we find that $|\lam_{345}|$ needs to be small, $\lesssim\,0.14$ for $m_H \approx 100$ \GeV{} and even tighter bounds for lighter $m_H$. However, there also exists a small mass window, $m_H\,\in\,[m_h/2; 63\,\GeV]$ which allows for values of $\lam_{345}$ up to our theoretical limit of $4 \pi$. As can be seen in the right of Fig.~\ref{fig:dmconst}, this region predicts particularly small values of $\Omega$ and therefore avoids both relic density and direct detection constraints, see Eq.~(\ref{eq:xenonlim}).  When the intermediate SM-like Higgs boson $h$ is on shell, the annihilation cross section  $H\,H\,\rightarrow\,b\,\bar{b}$ is enhanced and results in a considerably smaller dark matter relic density (see also discussion in \cite{Ilnicka:2015jba,Belyaev:2016lok}).

\subsubsection*{Combination}
\begin{figure}[tb!]
  \begin{center}
    \begin{minipage}{0.48\textwidth}
      \includegraphics[width=\textwidth]{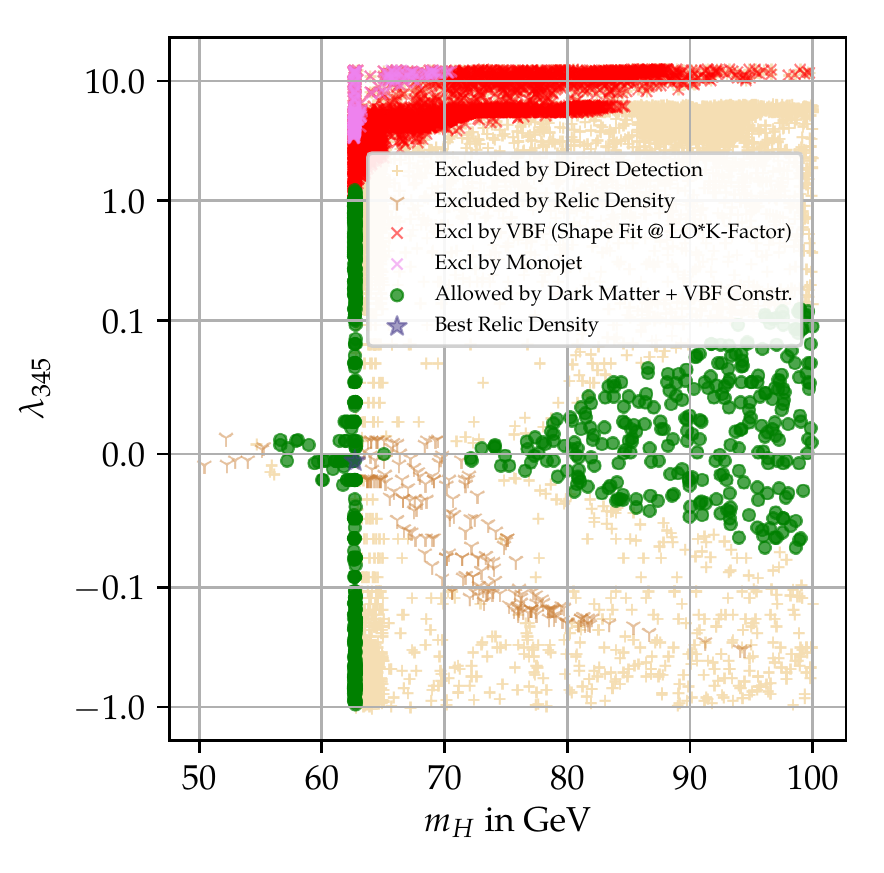}
    \end{minipage}
    \begin{minipage}{0.48\textwidth}
      \includegraphics[width=\textwidth]{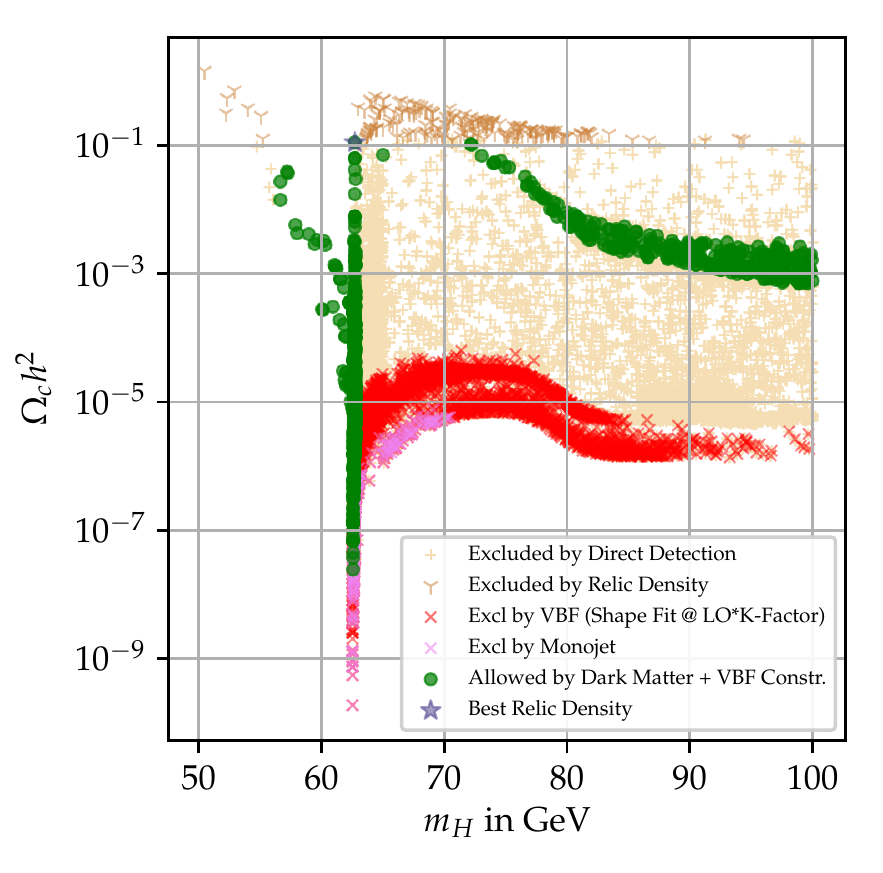}
    \end{minipage}
  \end{center}  \caption{\label{fig:finres} Parameter space after including all constraints (see also explanations below Fig.~\ref{fig:dmconst})}
\end{figure}
Finally, in Fig.~\ref{fig:finres}, we show the allowed and excluded parameters $m_H$ and $\lam_{345}$, as well as $\Omega_c h^2$, after all the above constraints are taken into account. As can be seen, collider results start to close the annihilation window, $m_H \approx m_h/2$, which could bypass direct detection constraints by significantly reducing the predicted relic density $\Omega_c h^2$. Here, only collider searches can {put bounds on} values of $\lam_{345}$ above 1. However, this only constrains points with very small $\Omega_c h^2$; therefore still a large number of points in this kinematic window remain allowed, including our ``Best Relic Density'' point discussed above. Moreover, for values of $m_H$ significantly larger than $m_h/2$, collider limits may yield important bounds. However in the IDM we find that these are always already excluded by direct detection limits.

It must be noted, though, that if the lightest IDM scalar $H$ couples to an extended dark sector and in fact decays to the actual, lighter dark matter candidate, relic density and direct detection constraints can change significantly while the above collider bounds are typically unaffected if $H$ has further invisible decays (see e.g.\ Ref.~\cite{Ibarra:2016dlb} in the context of the so-called ``radiative seesaw model'' which extends the IDM with an additional Majorana neutrino dark matter candidate). Therefore, even though in the pure IDM collider limits seem to hardly provide additional sensitivity compared to direct detection limits, they still constitute  an important analysis channel complementary to dark matter findings.

  \section{Null Results from other recast channels}

  The above VBF and monojet analyses focus on the dark matter candidate $H$ and thus are largely independent of the masses $m_A, m_{H^\pm}$ of the other two scalar particles and their decay rates. However, within our scan we {only considered} dark masses $\leq\,500\,\GeV$. Thus the question may arise if any of the many other BSM searches performed by ATLAS and CMS could result in additional, stronger constraints than the one considered. 

\begin{figure}[tb!]
    \begin{center}
      \begin{minipage}{0.32\textwidth}
        \includegraphics[width=\textwidth]{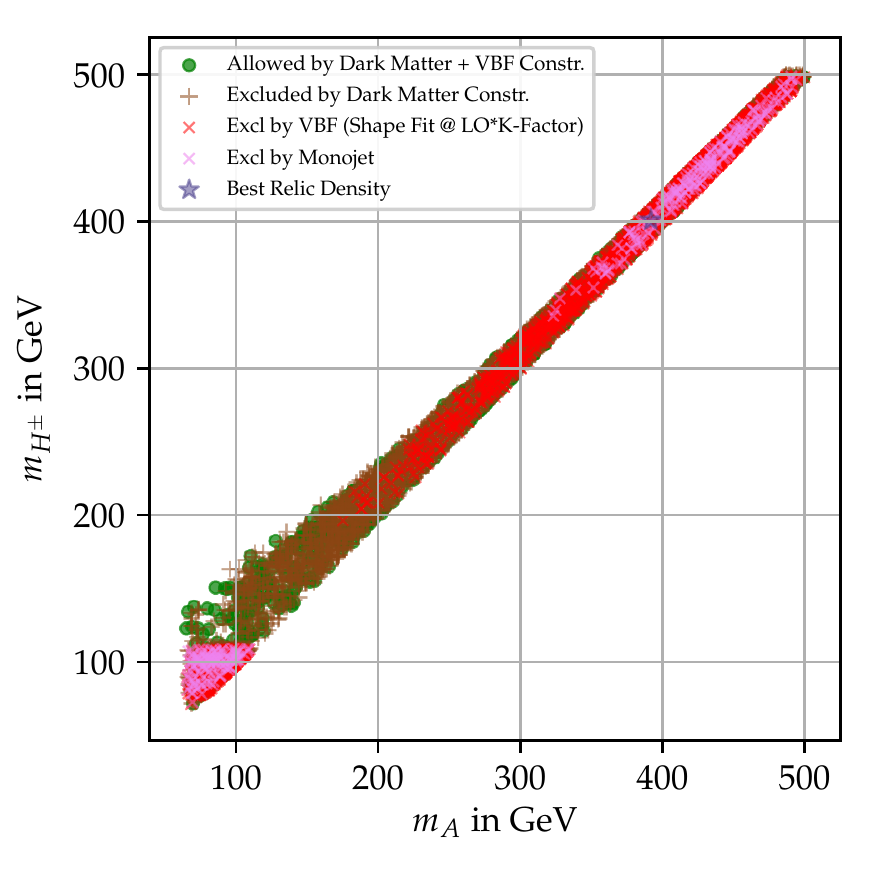}
      \end{minipage}
      \begin{minipage}{0.32\textwidth}
        \includegraphics[width=\textwidth]{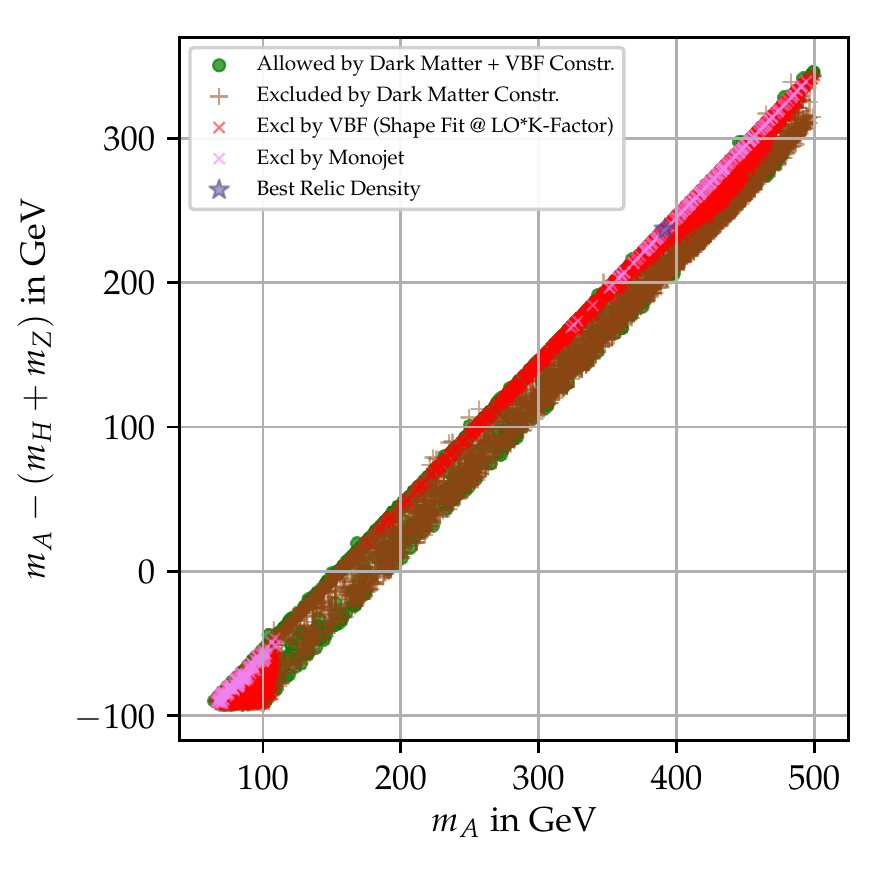}
      \end{minipage}
      \begin{minipage}{0.32\textwidth}
        \includegraphics[width=\textwidth]{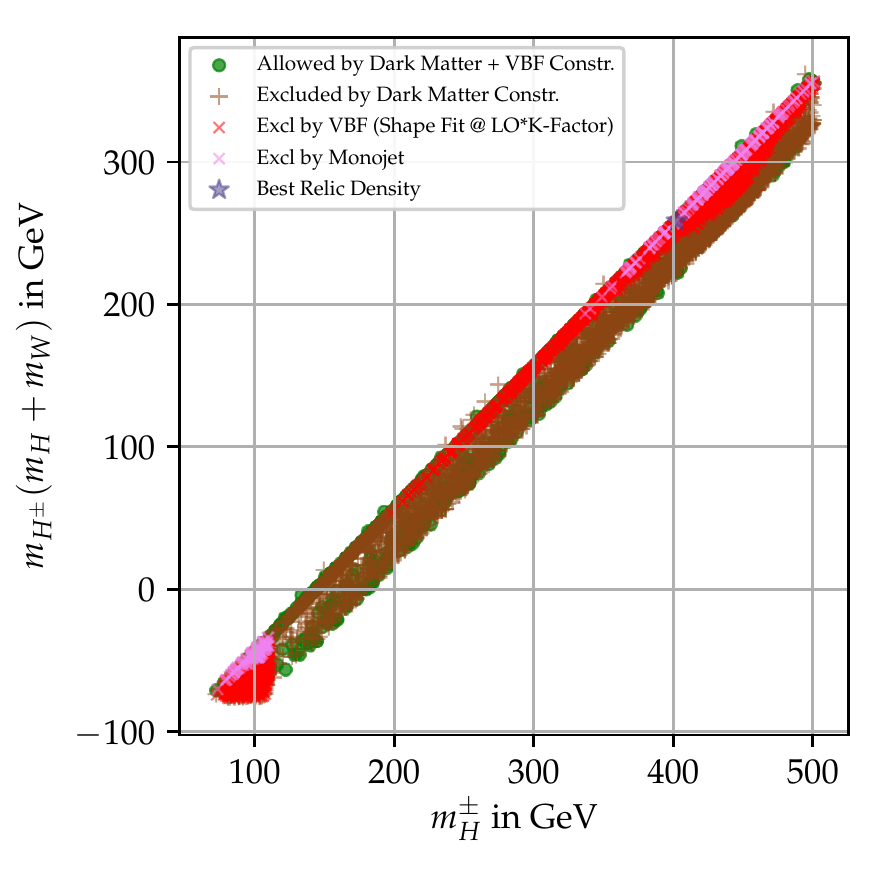}
      \end{minipage}
    \end{center}
    \caption{\label{fig:mAmHp} Allowed combinations of the scalar masses $m_{H^{\pm}}$, $m_A$ and $m_H$ which are relevant for collider analysis of the channels $p p \rightarrow H A$ and $p p \rightarrow H^\pm A$.}
  \end{figure}

In Fig.~\ref{fig:mAmHp} we display the masses $m_{H^\pm}, m_A$ for all points that are allowed by our previous scan. Similar to findings in Refs.~\cite{Ilnicka:2015jba,Kalinowski:2018ylg}, we observe a relatively strong mass degeneracy of these two heavier dark scalars. We also show the corresponding mass differences $m_{A} - (m_H + m_Z)$ and $m_{H^\pm} - (m_H + m_{W^\pm}$) of the allowed points. This quantity can be used to roughly estimate the kinematics of the expected decays for the heavier scalars $A$ and $H^\pm$. For mass differences larger/smaller than 0, we expect on/off-shell decays into gauge bosons, e.g.\ $A \rightarrow H Z^{(*)}$, with 100\% branching ratio due to the absence of any other lighter $D$-odd particles. We focus on the leptonic decay modes of the gauge bosons as within the analyses we consider,  hadronic modes are typically harder to distinguish from QCD background.

The larger the mass difference, the more energy is expected to be passed on to the daughter particles. A high-momentum $H$ in the final state is expected to produce missing transverse momentum (MET) in the event, a key observable for BSM signals. However, as can be seen in Fig.~\ref{fig:mAmHp}, viable IDM points only predict large mass differences for parameter points if also the absolute masses of $m_A$ and $m_{H^\pm}$ are increased.  Points with a large mass splitting  and good final state efficiency therefore in turn suffer from respectively smaller expected LHC production cross sections. Therefore, a dedicated Monte Carlo recast procedure is necessary in order to identify which points are subject to constraints from direct LHC searches.

 Fortunately, \checkmate{} bears the great advantage of being capable of quickly testing many such analyses simultaneously. We hence used it to perform a more inclusive scan of other potentially relevant final states. To be more precise, we considered the two body final states 
  \begin{align}
    p p \rightarrow H A,\ H H^\pm,\ A A,\ A H^\pm \text{ and } H^+ H^-.
  \end{align}

  We simulated all above processes in \texttt{MG5\_aMC@NLO}, including a full consideration of the 2- and 3-body decays of $A$ and $H^\pm$ into $H$ and a set of Standard Model particles. These events are subsequently tested against all 13 TeV analyses implemented in \checkmate{}  --- a list is given in appendix \ref{app:checkmateanal}. 

  It turns out that none of our $> 10,000$ considered parameter tuples appear to be excluded by any search other than the already considered VBF and monojet channels. 

  Note that though we inclusively test all possible final states, the highest sensitivity is expected from leptonic final states, i.e.\
  \begin{align}
    p p \rightarrow &A H, A \rightarrow Z_{\text{lep}}^{(*)} H \label{eq:pro1}\\
    p p \rightarrow &A H^\pm, A \rightarrow Z_{\text{lep}}^{(*)} H, H^\pm \rightarrow W^{\pm}_{\text{lep}} H. \label{eq:pro2}
  \end{align}
  The first signature is covered by Ref.~\cite{Aaboud:2017bja} which searches for final state with invisible particles produced in association with a leptonically decaying $Z$-boson.\footnote{Note that this final state has been analysed before in Ref.~\cite{Belanger:2015kga} using Run 1 dilepton final states. However, the parameter regions they consider are excluded after applying constraints from dark matter relic density and the invisible width of the SM Higgs boson.}  We refer to this analysis as ``$2 \ell$'' in the following. In contrast, the second example signature is covered\footnote{{Note that \texttt{Checkmate}, and therefore also our analysis, makes use of preliminary results in Ref.~\cite{ATLAS-CONF-2017-039} which were subsequently updated by a full publication in Ref.~\cite{Aaboud:2018jiw}. However, the published results are identical to those in the preliminary conference note.}} by Ref.~\cite{ATLAS-CONF-2017-039} --- for short ``$3 \ell$'' in the following text --- which looks for various leptonic (and hadronic) final states in supersymmetric electroweakino production, i.e.\ $\tilde \chi^+_1 \tilde \chi^-_1$ and $\tilde \chi^\pm_1 \tilde \chi^0_2$.  The expected final state for mixed chargino-neutralino production is experimentally identical to the aforementioned $A H^\pm$ decay chain and thus may be used to constrain the IDM.

\begin{figure}[tb!]
    \includegraphics[width=0.48\textwidth]{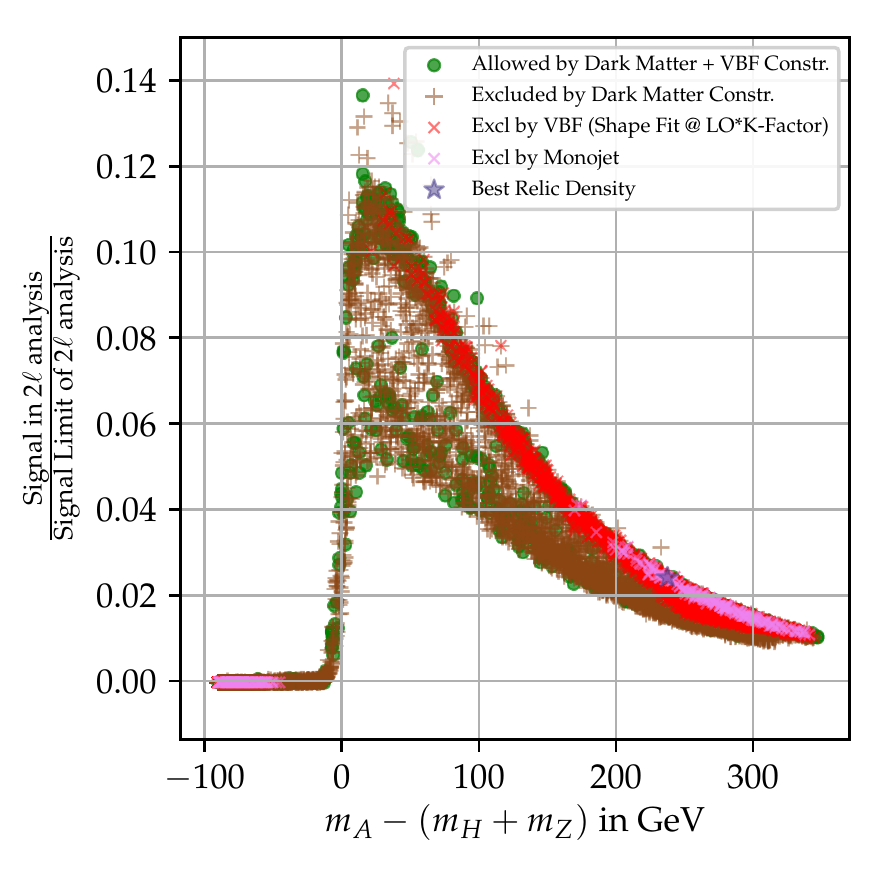}
    \includegraphics[width=0.48\textwidth]{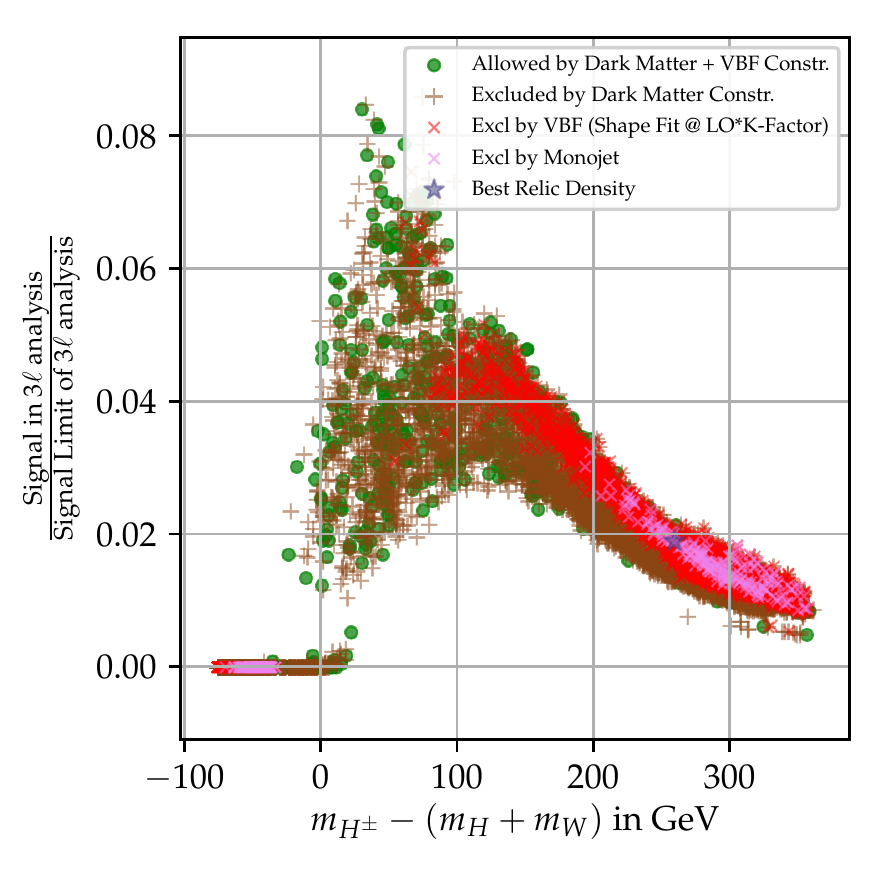}
    \caption{Results for our LHC reinterpretation of SUSY electroweakino results on the masses of the IDM. The $x$-axis shows mass differences which are strongly correlated to the MET distribution in the final state. The $y$-axis denotes the $r$-value, defined as the ratio of the signal prediction divided by the 95 \% confidence limit on the signal.}
    \label{fig:checkmatew}
  \end{figure}

  To understand the reason for the non-sensitivity of current electroweakino searches, we show our results for the $AH$-channel and the $H^\pm A$ channel in Fig.~\ref{fig:checkmatew}. In each subplot, we show the respective $r$-value of the analysis, defined as the ratio of the signal predicted by \texttt{CheckMATE} for the most sensitive signal region and the model-independent upper limit on a signal in this signal region.
  Most importantly, $r$ scales with the predicted signal cross section and a value of $r \geq 1$ can be interpreted as a model point excluded at 95\% confidence level.

  For the $x$-axis, we respectively show the mass difference of a heavy inert scalar, $A$ or $H^\pm$, and the summed masses of the two particles it decays into, e.g.\ $m_A - (m_Z + m_H)$ for $A \rightarrow Z H$. As explained before, the mass difference provides an estimate for the typical energy given to the leptons and to the dark matter candidate $H$ in the form of MET. 

  As can be seen from the figures, there is no 1:1 correspondence between the aforementioned mass difference and the model exclusion. This is obvious since the limit also depends on the absolute mass scales which for a given mass difference can change within $\pm 50$ GeV, c.f.~Fig.~\ref{fig:mAmHp}. However, one observes an overall rise-and-fall of the sensitivity and a global maximum near $m_A - (m_H + m_Z) \approx 50$ \GeV{} and $m_{H^\pm} - (m_H + m_{W^\pm}) \approx 125$ \GeV. This structure can be explained from our discussion at the beginning of this section: The larger the mass difference, the higher the expected amount of lepton $p_T$ and MET in the final state becomes and so the overall signal efficiency increases.  However, in order to obtain higher mass differences, electroweak precision constraints require larger masses for $A$ and $H^\pm$ and thus generally predicts smaller cross sections for viable IDM realisations. Hence, large mass differences simultaneously increase the final state efficiency and decrease the expected cross section. For the $2 \ell$ analysis, this results in a peak at a mass difference of approximately $30$ GeV which is  related to the minimum $p_T$ cut on the signal leptons and the MET requirement of this analysis. For the $3 \ell$ analysis, no overall peak can be determined as the final state consists of two separate decay chains whose kinematic configurations simultaneously depend on $m_A$ and $m_{H^\pm}$. Still, a similar behaviour can be observed.

  \begin{figure}[tb!]
    \includegraphics[width=0.48\textwidth]{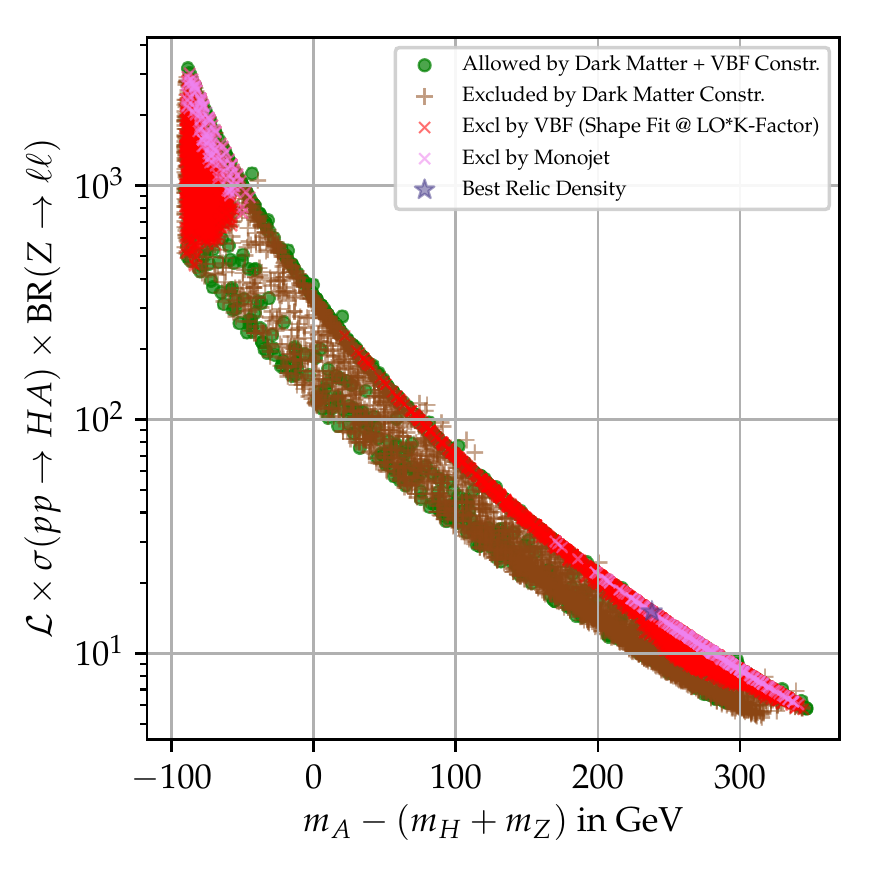}
    \includegraphics[width=0.48\textwidth]{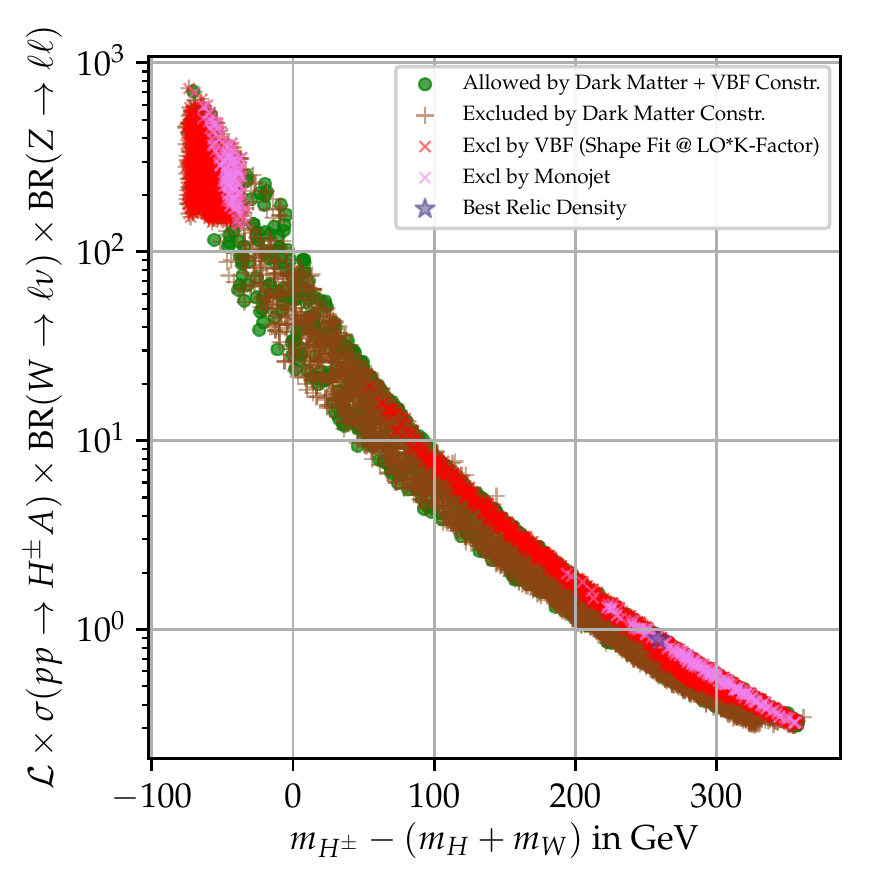}
    \caption{Predicted number of events in the $2\ell$/$3\ell$ channels.  The $x$-axis shows mass differences as in Fig.~\ref{fig:checkmatew}. The $y$-axis denotes the product of integrated luminosity, total production cross section and leptonic branching ratio of the expected gauge boson(s) in the final state.}
    \label{fig:checkmatew2}
  \end{figure}
  
  However, as can be seen, the peak values for both analyses still only predict at most 15 \% of the required number of events for the analyses to be sensitive to the signal. Hence, we conclude that electroweakino searches are currently not sensitive to the IDM and from a statistical point of view, this may only change in the high luminosity limit of LHC 14. Still, the presented analysis only shows reinterpreted results motivated from different signal models and therefore not necessarily optimised towards the IDM. It may therefore be possible that a collider search specifically targeting the IDM may improve upon the results determined here via reinterpretation.

  As an example, Fig.~\ref{fig:checkmatew2} illustrates how many events with leptonic final states are respectively expected from the processes in Eqs.~(\ref{eq:pro1}), (\ref{eq:pro2}), \emph{without} applying any event selection cuts. Note that for a signal to be observable e.g.\ in the $2\ell$ analyses one requires at least 200 events \emph{after} requiring the missing transverse momentum to be at least 90 GeV. It becomes apparent from Fig.~\ref{fig:checkmatew2} that a considerably softer cut on MET would significantly increase the number of expected signal events after cuts within the IDM. A full sensitivity study would however require the re-evaluation of SM background after modifying cuts.  Such an analysis would provide important complementary information since, as can be seen from the different categories shown in Fig.~\ref{fig:checkmatew}, many points in the peak region of the direct search are neither excluded by the VBF channel nor by dark matter direct detection.

  \section{Conclusions}
  In this paper, we have considered the Inert Doublet Model, a two Higgs doublet model with a discrete $Z_2$ symmetry containing a scalar dark matter candidate. We have included all current theoretical and experimental collider constraints on this model as discussed in Ref.~\cite{Kalinowski:2018ylg}. Concentrating on the region where $m_H\,\leq\,100\,\GeV$, we have investigated limits on the models' parameter space from a recast of recent LHC search where the invisibly decaying SM Higgs is produced either in vector boson fusion, Ref~\cite{Sirunyan:2018owy}, or in association with a hard jet, Ref.~\cite{ATLAS-CONF-2017-060}. For this, we have implemented the above searches in the collider phenomenology tool \texttt{CheckMATE} and tested their sensitivity compared to constraints from dark matter and direct detection. 

We observe that the VBF channel outperforms the monojet analysis and is sensitive to a large fraction of IDM parameter space and a proper recast of this analysis results in important bounds on the IDM model. Our search can significantly constrain a specific window in parameter space with dark matter masses $\sim\,62-63\,\GeV$ which evades dark matter limits due to an enhanced annihilation rate and leads to a significantly reduced relic abundance. This softens constraints from direct detection experiments like \texttt{XENON1T}. For larger masses, the VBF channel still provides relevant bounds which however do not improve direct detection limits within the pure IDM. The latter, however, could be avoided by coupling the lightest IDM scalar to a lighter dark matter sector which would have nearly no consequence for our presented collider analysis. 

As no direct search for IDM scalars exist, we further reinterpret searches for BSM particles with the same experimental signature and conclude that these do not put further constraints on the IDM. We trace this back to the effect that either the cross section is too small or the mass splitting is not large enough to predict sufficiently high-energetic final state particles. In this context, it might be interesting to pursue whether a dedicated search for the inert scalars could enhance the expected LHC sensitivity and eventually provide complementary information to the VBF channel and dark matter direct detection, especially about the other scalar masses of the dark sector.

\section*{Acknowledgements}
This research was supported in parts by the National Science Centre, Poland, the HARMONIA project under contract UMO-2015/18/M/ST2/00518 (2016-2019), the National Science Foundation under Grant No. 1519045, by Michigan State University through computational resources provided by the Institute for Cyber-Enabled Research, by grant K 125105 of the National Research, Development and Innovation Fund in Hungary, and by the European Union through the European Regional Development Fund - the Competitiveness and Cohesion Operational Programme (KK.01.1.1.06). TR also thanks the DESY Theory group for their hospitality while parts of this work were completed. DD acknowledges funding and support from the Collaborative Research Unit (SFB) 676 of the Deutsche Forschungsgemeinschaft (DFG), project B1.

  \begin{appendix}
    \section{IDM Feynman rules and other relations}\label{app:fr}
    The parameters $m_{22}^2,\,\lam_3,\,\lam_4,\,\lam_5$ {can be re-expressed} in terms of our input parameters: 
    \begin{eqnarray*}
      &&m_{22}^2\,=\,\lam_{345}\,v^2-2\,m_H^2,\\
      &&\lam_3\,=\,\lam_{345}-\frac{2}{v^2}\,\lb m_H^2-m_{H^\pm}^2 \rb,\,\lam_4\,=\,\frac{m_A^2+m_H^2-2\,m_{H^\pm}^2}{v^2},\\
&&\lam_5\,=\,\frac{m_H^2-m_A^2}{v^2}.
    \end{eqnarray*}

For completeness, we list the relevant Feynman rules of the IDM scalars in Tables \ref{tab:scal},\ref{tab:quar} and \ref{tab:gb}, omitting Goldstone modes as we are working in the unitary gauge at tree level. Note that the second, inert doublet neither participates in electroweak symmetry breaking nor in the generation of fermion masses. Hence, the couplings of the SM-like Higgs {$h$} to electroweak gauge bosons as well as fermions are given by their SM values, {see e.g.}  \cite{Bohm:2001yx}, {with {the} convention $g_{h W^+_\mu W^-_\nu}=ie^2 v/2 {s_W}^2g_{\mu\nu}$ }.

\begin{table}[h!]
  \begin{minipage}[T]{0.3\textwidth}
    \begin{tabular}{l|c}
\hline \hline
      {vertex}& coupling\\ \hline
      $hHH$&$\lam_{345}\,v$\\
      $hAA$&$\bar{\lam}_{345}\,v$\\
      $hhh$&$3\,\lam_1\,v$\\
      $h\,H^+\,H^-$&$\lam_3\,v$\\
      \hline
\hline
    \end{tabular}
\caption{Triple scalar vertices}
\label{tab:scal}
  \end{minipage}
  \begin{minipage}[T]{0.3\textwidth}
    \begin{tabular}{l|c}
\hline \hline
      {vertex}& coupling\\ \hline
      $hhhh$&$3\,\lam_1$\\
      $H^+\,H^+\,H^-\,H^-$&$2\,\lam_2$ \\
      $HHAA$&$\lam_2$\\
      $HHHH$&$3\,\lam_2$\\
      $AAAA$&$3\,\lam_2$\\
      $H^+H^-\,AA$&$\lam_2$\\
      $H^+H^-HH$&$\lam_2$\\
      $hhH^+H^-$&$\lam_3$\\
      $hhHH$&$\lam_{345}$\\
      $hhAA$&$\bar{\lam}_{345}$ \\
\hline \hline
    \end{tabular}
\caption{Quartic scalar vertices}
\label{tab:quar}
  \end{minipage}
  \begin{minipage}[t]{0.3\textwidth}
    \begin{tabular}{l|c}
\hline \hline
      {vertex}& coupling\\ \hline
      $H^-\,H^+\,\gamma$&$i\,e$\\
      $H^-\,H^+\,Z$&$i\,\frac{g}{2}\,\frac{\cos\,(2\theta_W)}{\cos\theta_W}$\\
      $H\,H^\pm\,W^\mp$&$\mp\,i\,\frac{g}{2}$\\
      $A\,H^\mp\,W^\pm$&$-\frac{g}{2}$\\
      $H\,A\,Z$&$-\frac{g}{2\cos\theta_W}$ \\
\hline \hline
    \end{tabular}
\caption{Gauge-scalar vertices}
\label{tab:gb}
  \end{minipage}
\end{table}

    \section{List of Applied CheckMATE Analyses}
    \label{app:checkmateanal}
    \nopagebreak
    \newcommand{\etmiss}{$\slashed{E}_T$}
    \begin{table*}
      \scriptsize
      \setlength{\tabcolsep}{1.1pt}
      \def\arraystretch{0.70}
      \begin{tabularx}{\textwidth}{lXlll}
        \hline
        \hline
        \checkmate{} identifier \hspace{0.5cm} \phantom{0}&  Search designed for&  \#SR & $L_{\text{int}}$ & Ref.\\
        \hline
        \texttt{atlas\_1602\_09058} & Supersymmetry in final states with jets and two SS leptons or 3 leptons & 4 & 3.2 & \cite{Aad:2016tuk}\\   
        \texttt{atlas\_1604\_01306} & New phenomena in events with a photon and \etmiss{} & 1 & 3.2 & \cite{Aaboud:2016uro}  \\   
        \texttt{atlas\_1604\_07773} & New phenomena in final states with an energetic jet and large \etmiss{} & 13 & 3.2 & \cite{Aaboud:2016tnv}  \\   
        \texttt{atlas\_1605\_03814} & $\tilde q$ and $\tilde g$ in final states with jets and \etmiss{} & 7 & 3.2  & \cite{Aaboud:2016zdn} \\   
        \texttt{atlas\_1605\_04285} & Gluinos in events with an isolated lepton, jets and \etmiss{} & 7 & 3.3  & \cite{Aad:2016qqk} \\   
        \texttt{atlas\_1605\_09318} & Pair production of $\tilde g$ decaying via $\tilde t$ or $\tilde b$ in events with $b$-jets and \etmiss{} & 8 & 3.3 & \cite{Aad:2016eki}   \\   
        \texttt{atlas\_1606\_03903} & $\tilde t$ in final states with one isolated lepton, jets and \etmiss{} & 3 & 3.2  & \cite{Aaboud:2016lwz} \\   
        \texttt{atlas\_1609\_01599} & Measurement of $ttV$ cross sections in multilepton final states & 9 & 3.2  & \cite{Aaboud:2016xve} \\   
        \texttt{atlas\_conf\_2015\_082} & Supersymmety in events with leptonically decaying $Z$, jets and \etmiss{} & 1 & 3.2  & \cite{ATLAS-CONF-2015-082} \\   
        \texttt{atlas\_conf\_2016\_013} & Vector-like $t$ pairs or 4 $t$ in final states with leptons and jets & 10 & 3.2 & \cite{ATLAS-CONF-2016-013}   \\   
        \texttt{atlas\_conf\_2016\_050} & $\tilde t$ in final states with one isolated lepton, jets and \etmiss{} & 5 & 13.3 & \cite{ATLAS-CONF-2016-050} \\   
        \texttt{atlas\_conf\_2016\_054} & $\tilde q$, $\tilde g$ in events with an isolated lepton, jets and \etmiss{} & 10 & 14.8 & \cite{ATLAS-CONF-2016-054} \\   
        \texttt{atlas\_conf\_2016\_076} & Direct $\tilde t$ pair production and DM production in final states with 2$\ell$ & 6 & 13.3 & \cite{ATLAS-CONF-2016-076}  \\   
        \texttt{atlas\_conf\_2016\_078} & Further searches for $\tilde q$ and $\tilde g$ in final states with jets and \etmiss{} & 13 & 13.3 & \cite{ATLAS-CONF-2016-078}  \\   
        \texttt{atlas\_conf\_2016\_096} & Supersymmetry in events with 2$\ell$ or $3\ell$ and \etmiss{} & 8 & 13.3 & \cite{ATLAS-CONF-2016-096} \\   
        \texttt{atlas\_conf\_2017\_022} & $\tilde q$, $\tilde g$ in final states with jets and \etmiss{} & 24 & 36.1 & \cite{ATLAS-CONF-2017-022}\\
        \textbf{\texttt{atlas\_conf\_2017\_039}} & \textbf{Electroweakino production in final states with 2 or 3 leptons} & 37 & 36.1 & \cite{ATLAS-CONF-2017-039}  \\   
        \textbf{\texttt{atlas\_conf\_2017\_040}} & \textbf{Dark Matter or invisibly decaying $h$, produced in associated with a $Z$} & 2 & 36.1  & \cite{ATLAS-CONF-2017-040}\\ 
        \textbf{\texttt{atlas\_conf\_2017\_060}} & \textbf{New phenomena in final states with an energetic jet and large \etmiss{}} & 13 & 36.1 & \cite{Aaboud:2017phn}  \\     
        \texttt{cms\_pas\_sus\_15\_011} & New physics in final states with an OSSF lepton pair, jets and \etmiss{} & 47 & 2.2 &  \cite{CMS-PAS-SUS-15-011}   \\
        \textbf{\texttt{cms\_pas\_hig\_17\_023}} & \textbf{Search for invisible decays of $h$ produced through VBF}& 10 & 36.1 &  \cite{Sirunyan:2018owy}   \\
        \hline
        \hline
      \end{tabularx}
      \caption{Full list of all $\sqrt{s} = 13$~ \TeV{} \checkmate{} analyses used for this
        study. Entries in boldface are relevant for the model studied in this work and are discussed in the main text. The column labelled \#SR yields the number of signal
        regions. Entries for the integrated luminosities $L_{\text{int}}$
        are given in fb$^{-1}$.} 
      \label{tab:app:analyses}
    \end{table*}
    Table~\ref{tab:app:analyses} gives the full list of used \checkmate{}
      analyses with a centre-of-mass energy of $\sqrt{s}$ = 13~\TeV. The first column shows the \checkmate{} idenitifer, the
      second the purpose for which the analysis was designed for. The last
      three columns show the number of signal regions in the corresponding
      analysis (marked \#SR), the integrated luminosity for that analysis
      and the reference to the publication or conference notes from the
      experimental collaborations. We mark all analyses discussed in our main discussion in boldface. Note that \texttt{Checkmate} regularly implements preliminary results published as conference notes by the experimental LHC collaborations and use the corresponding \texttt{conf-note} identifiers. Often, these are published at a later stage by the collaborations without any changes to analysis procedure or results. More details on the individual analyses can be found in their respective references 
      and corresponding validation material, if not provided in this work,  can be found on
      \url{http://checkmate.hepforge.org}.

  \end{appendix}

%\bibliography{lit,lita}
%\input{main_3_rev_proofs.bbl}

\end{document}